\documentclass[prb,aps,twocolumn,amsmath,amssymb]{revtex4-1}
\usepackage{graphicx}
\usepackage{dcolumn}
\usepackage{bm}
\usepackage{color}
\usepackage{float}
\usepackage{wrapfig}
\usepackage{comment}
\usepackage{soul}
\allowdisplaybreaks[1]

\usepackage{graphicx}

\usepackage{braket}
\usepackage{cases}
\usepackage{mathrsfs}

\usepackage{here}
\usepackage{float}

\usepackage{tablefootnote}
\usepackage{footnote}

\begin{document}
\title{Dzyaloshinskii-Moriya interaction in strongly spin-orbit-coupled systems: \\
General formula and application to topological and Rashba materials}
\date{\today}
\author{Yuto Hayakawa, Yusuke Imai, and Hiroshi Kohno}
\affiliation{Department of Physics, Nagoya University, Nagoya 464-8602, Japan}

\begin{abstract}
  We theoretically study the Dzyaloshinskii-Moriya interaction (DMI)   
mediated by band electrons with strong spin-orbit coupling (SOC).  
  We first derive a general formula for the coefficient ${\bm D}_i$ of the DMI in free energy
in terms of Green's functions, and examine its variations in relation to physical quantities. 
 In general, the DMI coefficient can vary depending on physical quantities, 
i.e., whether one is looking at equilibrium spin structure (${\bm D}_i$) or spin-wave dispersion (${\bm D}_i^{(2)}$),  
and the obtained formula helps to elucidate their relations.  
 By explicit evaluations for a magnetic topological insulator and a Rashba ferromagnet with perpendicular magnetization, 
we observe ${\bm D}_i^{(2)} \ne {\bm D}_i$ in general. 
 In the latter model, or more generally, when the magnetization and the spin-orbit field are mutually orthogonal, 
${\bm D}_i$ is exactly related to the equilibrium spin current for arbitrary strength of SOC, 
generalizing the similar relation for systems with weak SOC. 
 Among various systems with strong SOC, magnetic Weyl semimetals are special in that ${\bm D}_i^{(2)} = {\bm D}_i$, 
and in fact, the DMI in this system arises as the chiral anomaly. 
\end{abstract}

\maketitle

\section{Introduction}

 The Dzyaloshinskii-Moriya interaction (DMI) \cite{Dzyaloshinsky,Moriya} is a unique exchange 
interaction 
that favors a mutual twist of spins in solids and molecules. 
 In ferromagnetic materials, in which long-wavelength magnetization structure is of primary interest, 
the DMI is described by an effective free energy 
(\lq\lq Dzyaloshinskii-Moriya (DM) free energy'') \cite{Dzyaloshinsky1964} 
\begin{eqnarray}
 F_{\rm DM} &=&  \int d{\bm r}  \,  {\bm D}_i \cdot ({\bm n} \times \partial_i {\bm n}) , 
\label{eq:DMI}
\end{eqnarray}
where ${\bm n} = {\bm M}/|{\bm M}|$ is the unit vector of the magnetization ${\bm M}$, 
and ${\bm D}_i$ is a coefficient vector, called a DM vector. 
 Recently, the DMI has received renewed interest in spintronics in the context of 
fast domain wall motion,\cite{Thiaville2012,Chen2013,Ryu2013,Emori2013,Torrejon2014}  
chiral magnetic textures such as skyrmions \cite{Bogdanov1989,Rossler2006,Muhlbauer2009,Yu2010,Heinze2011,Nagaosa2013} 
and spin helices, \cite{Shirane1983,Ishimoto1986}  
nonreciprocal magnon propagation, \cite{Iguchi2015,Seki2016,Sato2016,Takagi2017} and so on. 
 These phenomena in turn provide physical principles of experimental quantification of DMI 
through measurements of domain-wall speed,\cite{Torrejon2014,24,25,26,27} 
length scale of magnetic texture,\cite{Schlotter2018,Bacani2019,Agrawal2019,Garlow2019,Zhou2020}  
and dispersion or nonreciprocity of spin waves.\cite{Zakeri2010,Korner2015,Lee2015a,Cho2015,Belmeguenai2015,Di2015,Nembach2015,Stashkevich2015,
Chaurasiya2016,Ma2016,Hrabec2017,Robinson2017,Ma2017,Kim2021}  
More detailed exposition can be found in a recent review paper.~\cite{review2023}

 Theoretically, the first microscopic derivation of DMI was given by Moriya, 
who considered insulating magnets with dominant superexchange (and less dominant direct exchange)  interaction, and treated the spin-orbit coupling (SOC) perturbatively.\cite{Moriya}  
 Recent interest lies also in metallic magnets, in which the DMI is expected to be mediated 
by conduction electrons (or band electrons in general).           
 Theoretical calculations of the DMI coefficients in such systems have been done in various ways 
by many authors.\cite{Katsnelson2010,Gayles2015,Koretsune2015,Wakatsuki2015,Kikuchi2016,Freimuth2017,Koretsune2018,Ado2018} 
 In the first-principles calculations, they evaluated the energy change associated 
with magnetization twists.\cite{Heide2008,Katsnelson2010,Gayles2015} 
 Others studied the magnon self-energy, which is given by the spin-spin correlation function.\cite{Koretsune2015,Wakatsuki2015} 
 Recently, 
 Kikuchi et al. pointed out that the DMI coefficient is given by the equilibrium spin current of electrons,  
and presented a picture of DMI in terms of the Doppler shift caused by the spin current.\cite{Kikuchi2016} 
 Most of the studies were done within first order in SOC, and are valid for systems with weak SOC. 
 However, many materials of recent interest possess extremely strong SOC, 
and a theoretical scheme applicable to such systems is needed.

 From a technical point of view, the DMI mediated by band electrons is obtained 
as an effective free energy for ${\bm M}$ by integrating out the electrons;\cite{Kataoka1984}  
the electrons influenced by the nonuniform ${\bm M}$ and SOC contribute to the free energy 
in a chirality (i.e., direction of twist) dependent way. 
 Therefore, the coefficient ${\bm D}_i$ generally depends on ${\bm M}$, 
and $F_{\rm DM}$ is a functional of ${\bm M} = {\bm M} ({\bm r})$ 
not only through ${\bm n} \times \partial_i {\bm n}$ but also through the coefficient 
${\bm D}_i ({\bm M})$.
 As a consequence, the DMI coefficient can be different for different physical quantities, 
such as the pitch of spin helix (determined by $F_{\rm DM}$) and the spin-wave dispersion 
(determined by the second-order variation of $F_{\rm DM}$). 
 There is an indication that the DMI coefficients calculated by magnetization twist and by spin susceptibility 
show appreciable disagreement \cite{Koretsune2018}. 
 Possible resolution may be found in the above observation.

 In this paper, we derive a general formula for the coefficient of the DMI 
mediated by band electrons with arbitrary strength of SOC, 
elucidate its general properties, and apply it to several systems with strong SOC. 
 The results are summarized as follows: 

(i) A general expression of ${\bm D}_i$ is derived in terms of Green's functions. 
 Only assumption made is that the magnetization couples to conduction electron spins ${\bm \sigma}$
through the $s$-$d$ type exchange interaction, $ {\bm M} \cdot {\bm \sigma}$. 
 The result is thus quite general and may be used for first-principles calculations. 

(ii) The DM vectors in free energy (${\bm D}_i$) and in spin-wave dispersion (${\bm D}_i^{(2)}$) 
or torque (${\bm D}_i^{(1)}$) are different in general.  
 In particular,  ${\bm D}_i^{(2)}$ and ${\bm D}_i$ are related via 
\begin{eqnarray}
 {\bm D}_i^{(2)} = \frac{M}{2} \frac{\partial  {\bm D}_i }{\partial M}  , 
\label{eq:DMI_relation_1}
\end{eqnarray}
or
\begin{align}
  {\bm D}_i ({\bm M}) &= \int_0^M \frac{2 dM' }{M'}  {\bm D}_i^{(2)} ({\bm M}') , 
\label{eq:DMI_relation_2}
\end{align} 
where $M$ is the magnitude of ${\bm M} = M {\bm n}$, and ${\bm M}' = M' {\bm n}$ 
has a common direction ${\bm n}$ with ${\bm M}$.

(iii) On the surface of a magnetic topological insulator (MTI), 
we find ${\bm D}_i \ne {\bm D}_i^{(2)}$ in the insulating state, 
whereas ${\bm D}_i = {\bm D}_i^{(2)}$ in the metallic state.

(iv) For a two-dimensional (2D) Rashba ferromagnet with parabolic dispersion 
and with perpendicular magnetization, 
${\bm D}_i$ and ${\bm D}_i^{(2)}$ are finite when only the lower band has Fermi surface (FS), 
and they are different, ${\bm D}_i \ne {\bm D}_i^{(2)}$. 

(v) For a magnetic Weyl semimetal, we find ${\bm D}_i = {\bm D}_i^{(2)}$. 
 This is probably an exceptional case among systems with strong SOC. 
 In fact, the DMI in this system is related to the chiral anomaly.

(vi) The spin-current formula \cite{Kikuchi2016} of ${\bm D}_i$, which holds at weak SOC, 
can be generalized to the case of strong SOC if the magnetization is perpendicular to the spin-orbit field.  
 In the generalized formula, ${\bm D}_i$ is given by the {\it change} of the equilibrium spin current 
of electrons caused by the coupling to magnetization.  
 A typical example is the Rashba ferromagnet with perpendicular magnetization.

 This paper is organized as follows. 
 In Sec.~II, we develop a general theory of the DMI mediated by band electrons. 
 In Sec.~III, we apply the results of Sec.~II to various systems, 
and calculate the DMI coefficient for MTI, Rashba ferromagnet, and magnetic Weyl semimetal. 
 Some discussions are given in Sec.~IV  on the relation between ${\bm D}_i$ and ${\bm D}_i^{(2)}$, 
and on the spin-current formula for ${\bm D}_i$. 
 The calculations outlined in Sec.~II are detailed in Appendix A (derivation of general formula) 
and Appendix B (calculation of free energy variations). 
 Case of weak SOC is examined in Appendix C, 
and the generalized spin-current formula is derived in Appendix D.

\section{General Formula}

 In this section, after describing a model, we derive a general formula for the DMI coefficient 
in free energy in terms of Green's functions. 
 We then study the DMI coefficients in several physical quantities by taking variational derivatives 
of the DM free energy.

\subsection{Microscopic model} 
\label{model}

 We consider band electrons described by the (nonmagnetic) Hamiltonian $H_0$ 
and the exchange coupling $H'$ to the magnetization ${\bm M}$, 
\begin{eqnarray}
 H_0 &=& \sum_{\bm k} c_{\bm k}^\dagger {\cal H}^0_{\bm k}  c_{\bm k} , 
\\
 H' &=&  \int d{\bm r} \, c^\dagger ({\bm r}) ({\bm M} ({\bm r}) \cdot {\bm \sigma} ) c ({\bm r})  
\\
&=&  \sum_{{\bm k}, {\bm q}}
       c_{{\bm k}+{\bm q}}^\dagger ( {\bm M}_{\bm q}  \cdot  {\bm \sigma} ) c_{\bm k}  . 
\label{eq:H_M}
\end{eqnarray}
 Here, $c_{\bm k}^\dagger = (\cdots , c_{n,{\bm k}}^\dagger, c_{n+1,{\bm k}}^\dagger, \cdots)$  
is a set of electron creation operators 
with $c_{n,{\bm k}}^\dagger = (c_{n,{\bm k} \uparrow}^\dagger, c_{n,{\bm k} \downarrow}^\dagger)$
for $n$th band. 
 In $H_0$, ${\cal H}^0_{\bm k}$ is a matrix, not necessarily diagonal, 
that defines (nonmagnetic) band structure with arbitrary type of SOC. 
 In $H'$, ${\bm M} ({\bm r}) = M {\bm n} ({\bm r})$ is the \lq\lq magnetization'', 
and ${\bm M}_{\bm q}$ is its Fourier component. 
 Only assumption made in this paper is that ${\bm M}_{\bm q}$, when viewed as a matrix element 
between band (Bloch) electrons, does not depend on ${\bm k}$ and band indices.

 In a localized picture of ferromagnetism, ${\bm M}$ is given by 
\begin{align}
 {\bm M} &= - J_{\rm sd} {\bm S} = J_{\rm sd} \frac{v_0}{\gamma \hbar} {\bm M}_{\rm mag}  , 
\label{eq:M}
\end{align}
where ${\bm S}$ is a localized spin responsible for the ferromagnetic moment, 
$J_{\rm sd}$ is the \lq\lq s-d'' exchange coupling constant between ${\bm S}$ 
and conduction electrons. 
 In the last expression, ${\bm M}_{\rm mag}$ represents the true magnetization 
(as defined in electromagnetism), $v_0$ is the volume per localized spin ${\bm S}$, 
and the gyromagnetic constant is denoted by $- \gamma$. 
 For itinerant ferromagnets, ${\bm M}$ is a ferromagnetic order parameter (or simply a mean field) 
including the coupling constant. 
 The analyses in the present paper do not depend on which picture is taken, but we may occasionally 
use the terminology of the localized picture because of its conceptual simplicity.

 In the next section, we treat $H'$ as a perturbation (but to infinite order) to derive the DM free energy.  
 The unperturbed Green's function is  
\begin{align}
 G_0 ({\bm k},i\varepsilon_m) &=  ( i\varepsilon_m + \mu - {\cal H}_{\bm k}^0 )^{-1}  , 
\label{eq:G_0}
\end{align}
where $\varepsilon_m = (2m+1) \pi T$ is the Matsubara frequency, $T$ is the temperature, 
and $\mu$ is the chemical potential. 
 We occasionally set $\hbar = 1$.

\subsection{DM free energy}

 We first outline the derivation of the general formula of the DMI, or precisely, its coefficient, 
deferring the details to Appendix \ref{derivation}. 
 We start from the general expansion formula for free energy,\cite{AGD} 
\begin{widetext}
\begin{align}
 F &= \frac{1}{\beta} \sum_{n=1}^\infty \frac{(-)^{n-1}}{n!} 
           \int_0^\beta d\tau_1 \int_0^\beta d\tau_2 \cdots \int_0^\beta d\tau_n \, 
     \langle {\rm T}_\tau [ H'(\tau_1)  H'(\tau_2) \cdots  H'(\tau_n)  ]
      \rangle_{\rm conn} , 
\label{eq:F_start}
\end{align}
where $\tau_i$ are imaginary times, 
$H'(\tau) = e^{H_0 \tau} H' e^{-H_0 \tau}$ is the perturbation in the interaction representation, 
and  $\beta = T^{-1}$ is the inverse temperature. 
 It is understood that only the connected diagrams are considered. 
 Noting that $H'$ has off-diagonal matrix elements with respect to wave vector, 
we extract the external momentum ${\bm q}$ fed by ${\bm M}_{\bm q}$ in its first order. 
 Because of momentum conservation, we need to extract ${\bm M}_{-{\bm q}}$ as well. 
 After a diagrammatic consideration (see Appendix \ref{derivation}), we obtain 
\begin{align}
 F &=  \sum_{\bm q}  (- q_i) M_{\bm q}^\alpha  M_{-{\bm q}}^\beta   
         \sum_{n=2}^\infty \frac{1}{n} 
        T \sum_m  \sum_{\bm k}    \sum_{r=1}^{n-1}  
       {\rm tr} \biggl[\,  
        ({\cal H}'_{\bm 0} G_0)^{r-1}  
         \sigma^\alpha  \frac{\partial}{\partial k_i}  
         \left[ G_0 ({\cal H}'_{\bm 0}  G_0)^{n-r-1} \right] 
        (\sigma^\beta G_0)  \,\biggr]   . 
\end{align}
 Here, all the Green's functions have the same Matsubara frequency $\varepsilon_m$ 
and wave vector ${\bm k}$, 
and ${\cal H}'_{\bm 0} = {\bm M} \cdot {\bm \sigma}$ is defined with uniform ${\bm M}$. 
 The trace \lq\lq tr'' refers to the band index (including spin).
 Taking the $r$-sum, we have 
\begin{align}
 F &=  \sum_{\bm q}  (- q_i) M_{\bm q}^\alpha  M_{-{\bm q}}^\beta   
        T \sum_m  \sum_{\bm k}   
        \sum_{n=2}^\infty \frac{1}{n}  \frac{1}{(n-2)!} \left( \frac{\partial}{\partial \zeta} \right)^{n-2}  
         {\rm tr} \biggl[\,  G_\zeta 
         \sigma^\alpha  \frac{\partial G_\zeta}{\partial k_i}  
         \sigma^\beta   \,\biggr] \Biggr|_{\zeta=0}  , 
\end{align}
\end{widetext}
where $G_\zeta$ is defined by 
\begin{align}
 G_\zeta ({\bm k},i\varepsilon_m) 
&=  ( i\varepsilon_m + \mu - {\cal H}_{\bm k}^0 - \zeta {\bm M} \cdot {\bm \sigma})^{-1}  . 
\label{eq:G_zeta}
\end{align} 
 Note that an auxiliary parameter $\zeta$ has been introduced. 
 If the extra factor $1/n$ were absent, the sum over $n$ would yield 
$\sum_{n=2}^\infty \frac{1}{(n-2)!} \left( \frac{d}{d\zeta} \right)^{n-2} = e^{d/d\zeta} $, 
but its presence requires the use of integration with respect to $\zeta$. 
 See Appendix \ref{derivation} for details. 
 The result is 
\begin{align}
 F &=  \sum_{\bm q}  (- q_i) M_{\bm q}^\alpha  M_{-{\bm q}}^\beta  
      \int_0^1 \zeta d\zeta \,  T \sum_l  \sum_{\bm k}  
       {\rm tr} \biggl[\,   G_\zeta 
         \sigma^\alpha  \frac{\partial G_\zeta}{\partial k_i}  
         \sigma^\beta   \,\biggr]   . 
\end{align}
 Writing in real space, 
 $q_i M_{\bm q}^\alpha  M_{-{\bm q}}^\beta = i M^\alpha (\partial_i M^\beta ) 
   = i M^2 n^\alpha  (\partial_i n^\beta)$, 
we obtain Eq.~(\ref{eq:DMI}) with the coefficient, 
\begin{align}
 D_i^\alpha &= \frac{1}{2i} \varepsilon_{\alpha\beta\gamma}  M^2 
        \int_0^1 \zeta \, d\zeta \, T\sum_m \sum_{\bm k}  
        {\rm tr} \left[ \sigma^\beta \frac{\partial G_\zeta}{\partial k_i} \sigma^\gamma G_\zeta  \right] .  
\label{eq:result_zeta}
\end{align}
 By changing the integration variable to $M' \equiv \zeta M$, one may also write 
\begin{align}
 D_i^\alpha &= \frac{1}{2i} \varepsilon_{\alpha\beta\gamma}   
        \int_0^M M' dM' \, T\sum_m \sum_{\bm k}  
        {\rm tr} \left[ \sigma^\beta \frac{\partial G_{M'}}{\partial k_i} \sigma^\gamma G_{M'}  \right] , 
\label{eq:result}
\end{align}
where $G_M$ is defined by 
\begin{align}
 G_M ({\bm k},i\varepsilon_m) 
&=  ( i\varepsilon_m + \mu - {\cal H}_{\bm k}^0 - {\bm M} \cdot {\bm \sigma})^{-1}  . 
\label{eq:G_zeta}
\end{align}
 In the following, we may omit the prime on $M'$ (integration variable) 
when no confusion is anticipated.  
 Note that the integration is done with respect to the magnitude $|{\bm M}|$  
of the (uniform) magnetization vector while its direction ${\bm n} = {\bm M}/|{\bm M}|$ is kept fixed. 
 Thus, $D_i^\alpha$ generally depends on ${\bm n}$ as well.

\subsection{Variational derivatives of DM free energy} 
\label{Variations}

 As noted in the Introduction, precise forms (and magnitudes) of the DMI coefficient 
depend on physical quantities. 
 This is because they correspond to different-order variational derivatives of the DM free energy 
$F_{\rm DM}$,  
which we now calculate.

 In taking variations we note that ${\bm M}$ in ${\bm D}_i$ and that in 
${\bm M} \times \partial_i {\bm M}$ should be treated on equal footing. 
 Namely, ${\bm M}$ in ${\bm D}_i$ can also be considered spatially nonuniform. 
 (A theoretical justificatoin is given in Appendix \ref{functional}.) 
 This viewpoint is (conceptually) important in the following manipulations.

 To take variations of $F_{\rm DM}$ with respect to ${\bm M}$, 
it is convenient to introduce $\tilde {\bm D}_i$ by 
\begin{align}
  {\bm D}_i = M^2 \tilde {\bm D}_i  , 
\label{eq:D_tilde}
\end{align}
and write Eq.~(\ref{eq:DMI}) as 
\begin{align}
  F_{\rm DM} [{\bm M}] &= \int \tilde {\bm D}_i \cdot ( {\bm M} \times \partial_i {\bm M} ) \, d{\bm r} , 
\label{eq:d1_F}
\end{align}
where ${\bm M} = M {\bm n}$. 
 Let us consider the variation, 
$\delta F_{\rm DM} \equiv F_{\rm DM} [{\bm M} + \delta {\bm M}] - F_{\rm DM} [{\bm M}]$, 
by noting that the \lq unperturbed' ${\bm M}$ is not necessarily constant (spatially uniform) 
but can be textured. 
 Up to the second order in $\delta {\bm M}$, one finds 
$\delta F_{\rm DM} = \delta^{(1)} F_{\rm DM} + \delta^{(2a)} F_{\rm DM} + \delta^{(2b)} F_{\rm DM}$,  
\begin{align}
 \delta^{(1)} F_{\rm DM} 
&= \int \delta {\bm M} \cdot (\partial_i {\bm M} \times \tilde {\bm D}_i^{(1)}) \, d{\bm r} , 
\label{eq:d1_F}
\\
 \delta^{(2a)} F_{\rm DM} 
&= \int \delta {\bm M} \cdot (\partial_i \delta {\bm M} \times \tilde {\bm D}_i^{(2)}) \, d{\bm r} , 
\label{eq:d2a_F}
\\
 \delta^{(2b)} F_{\rm DM} 
&=  \int \delta {\bm M} \cdot (\partial_i {\bm M} \times \delta \tilde {\bm D}_i^{(2)}) \,d{\bm r} ,  
\label{eq:d2b_F}
\end{align}
where 
\begin{align}
 \tilde {\bm D}_i^{(1)}  =  2 \tilde {\bm D}_i^{(2)} &= {\bm \nabla}_{\bm M} \times (\tilde {\bm D}_i \times {\bm M}) , 
\label{eq:D^2_eq}
\end{align}
and 
$\delta \tilde {\bm D}_i^{(2)} = (\partial \tilde {\bm D}_i^{(2)} /\partial M^\alpha)\delta M^\alpha$.  
 See Appendix \ref{d^1F} and \ref{d^2F} for the calculation, 
and Eqs.~(\ref{eq:D1_def}), (\ref{app:eq:D^1}), (\ref{eq:D2_def}), (\ref{app:eq:D^2}), 
and (\ref{eq:D2b_def}) for the results.

  Equation (\ref{eq:D^2_eq}) holds generally, 
irrespective of the functional form of $\tilde {\bm D}_i ({\bm M})$. 
 If its form is restricted to the formula, Eq.~(\ref{eq:result_zeta}) or (\ref{eq:result}), 
one can proceed further. 
 Namely, one can show that 
\begin{align}
 {\rm div}_M \tilde {\bm D}_i  \equiv {\bm \nabla}_{\bm M} \cdot \tilde {\bm D}_i = 0 ,  
\label{eq:divD=0}
\end{align}
(see Appendix \ref{general_relations} and \ref{proof}), and Eq.~(\ref{eq:D^2_eq}) is simplified to 
\begin{align}
 \tilde {\bm D}_i^{(1)}  =  2 \tilde {\bm D}_i^{(2)} 
&= \frac{1}{M} \frac{\partial (M^2 \tilde {\bm D}_i)}{\partial M} .  
\label{eq:D^2_eq1}
\end{align}
 This relation shows that, if $\tilde {\bm D}_i$ ($= {\bm D}_i/M^2$) is independent of $M$, 
$\tilde {\bm D}_i^{(2)}$ and $\tilde {\bm D}_i$ coincide, 
$\tilde {\bm D}_i^{(2)} = \tilde {\bm D}_i \ (= \frac{1}{2} \tilde {\bm D}_i^{(1)})$. 
 Otherwise, they are different in general. 
 The results are summarized as  
\begin{align}
& \tilde D_i^\alpha 
= \frac{1}{2i} \varepsilon_{\alpha\beta\gamma}   
        \int_0^1 \zeta \, d\zeta \, T\sum_m \sum_{\bm k}  
        {\rm tr} \left[ \sigma^\beta \frac{\partial G_\zeta}{\partial k_i} \sigma^\gamma G_\zeta  \right] ,  
\label{eq:result_D_tilde}
\\
&  \tilde D_i^{(2),\alpha} 
= \frac{1}{4i} \varepsilon_{\alpha\beta\gamma}   
        \, T\sum_m \sum_{\bm k}  
       {\rm tr} \left[ \sigma^\beta \frac{\partial G_M}{\partial k_i} \sigma^\gamma G_M \right] . 
\label{eq:result_D_tilde_2}
\end{align}

\subsection{DMI and physical quantities} 
\label{phys}

 Here, we briefly look at how the free energy variations are related to physical quantities.

 First, the coefficient $\tilde {\bm D}_i$ itself determines the change in (free) energy 
when the magnetization is twisted. 
 This fact is often used in first-principles calculations of the DMI coefficient.\cite{Koretsune2018}  
 For a helical structure, 
\begin{align}
 {\bm n} ({\bm r}) 
&= {\bm e}_1 \cos ({\bm q} \cdot {\bm r}) \pm {\bm e}_2 \sin ({\bm q} \cdot {\bm r}) , 
\end{align}
where ${\bm e}_1$ and ${\bm e}_2$ are two orthogonal unit vectors, 
the DM free energy becomes 
\begin{align}
   F_{\rm DM} &= \pm q_i {\bm D}_i \cdot ({\bm e}_1 \times {\bm e}_2) , 
\label{eq:F_helix}
\end{align}
where ${\bm D}_i = M^2 \tilde {\bm D}_i$. 
 From the slope of this $q_i$-linear contribution, ${\bm D}_i$ can be extracted numerically. 
 In the equilibrium configuration, 
 ${\bm e}_1$, ${\bm e}_2$, and the direction of ${\bm q}$ are determined to minimize Eq.(\ref{eq:F_helix}). 
 To determine the magnitude of ${\bm q}$, we consider the exchange stiffness  
$\frac{1}{2} J (\partial_i {\bm n})^2$ as well, and minimize the sum. 
 This leads to 
\begin{align}
   q_i &= \mp {\bm D}_i \cdot ({\bm e}_1 \times {\bm e}_2) / J , 
\label{eq:q}
\end{align}
and ${\bm e}_1$ and ${\bm e}_2$ are determined to maximize
$\sum_i [{\bm D}_i \cdot ({\bm e}_1 \times {\bm e}_2]^2$. 
 The period of the helix is thus governed by ${\bm D}_i$ rather than ${\bm D}_i^{(2)}$.

 The first-order variation $\delta^{(1)} F_{\rm DM}$ defines a torque that affects the dynamics 
of ${\bm M}$,
\begin{align}
 {\bm M} \times \frac{\delta^{(1)} F_{\rm DM}}{\delta {\bm M}} 
&= {\bm M} \times (\partial_i {\bm M} \times \tilde {\bm D}_i^{(1)}) 
\nonumber \\
&= ({\bm n} \cdot {\bm D}_i^{(1)}) \, \partial_i {\bm n}   , 
\label{eq:d1_F_torque}
\end{align}
where ${\bm D}_i^{(1)} = M^2 \tilde {\bm D}_i^{(1)}$. 
 This torque works like a spin-transfer torque\cite{Tatara2008} driven by a spin current, 
$j_{{\rm s},i} \propto {\bm n} \cdot {\bm D}_i^{(1)}$.

 The second-order variations, $\delta^{(2a)} F_{\rm DM}$ and $\delta^{(2b)} F_{\rm DM}$, 
describe the dynamics of fluctuations $\delta {\bm M}$ around ${\bm M}$ (such as spin waves). 
 The first one, 
\begin{align}
 \delta^{(2a)} F_{\rm DM} 
&=  \int (\delta {\bm M} \times \partial_i \delta {\bm M}) \cdot \tilde {\bm D}_i^{(2)} \, d{\bm r} , 
\label{eq:d2a_F_2}
\end{align}
leads to the so-called spin-wave Doppler shift. 
 In fact, the spin-wave spectrum, $\omega_{\bm q}$, in the absence of the DMI is modified to   
\begin{align}
 \Omega_{\bm q} &= \omega_{\bm q} - 2 q_i ({\bm n} \cdot {\bm D}_i^{(2)}) , 
\label{eq:SW_Doppler}
\end{align}
where the second term, linear in the wave vector $q_i$, describes the effects of the DMI 
(${\bm D}_i^{(2)} = M^2 \tilde {\bm D}_i^{(2)}$). 
 Note that the Doppler shift is determined by the parallel component of ${\bm D}_i^{(2)}$,  \cite{Melcher1973,Kataoka1987,Udvardi2009} as in the spin-transfer torque mentioned above. 
 The second one, 
\begin{align}
 \delta^{(2b)} F_{\rm DM} 
&=  \int \delta M^\alpha \delta M^\beta 
\left( \partial_i {\bm M} \times \frac{\partial \tilde {\bm D}_i^{(2)}}{\partial M^\alpha} \right)^\beta \, d{\bm r} , 
\label{eq:d2b_F_2}
\end{align}
is finite only when ${\bm M}$ is nonuniform, $\partial_i {\bm M} \ne {\bm 0}$. 
 This may serve as a potential energy for magnons.

 The equality ${\bm D}_i^{(1)}  =  2 {\bm D}_i^{(2)}$ [Eq.~(\ref{eq:D^2_eq})] 
means that the torque and the spin waves are described by the same DMI coefficient. 
 On the other hand, the pitch of magnetic helix is described by the different DMI coefficient 
($ {\bm D}_i$). 

 Finally, we note that for $\tilde {\bm D}_i$ (or ${\bm D}_i$) only the component perpendicular 
to ${\bm M}$ is of physical relevance, as seen from Eqs.~(\ref{eq:DMI}) and (\ref{eq:q}).  
 In contrast, the parallel component is important for $\tilde {\bm D}_i^{(1)}$ (or ${\bm D}_i^{(1)}$) 
and $\tilde {\bm D}_i^{(2)}$ (or ${\bm D}_i^{(2)}$), 
as Eqs.~(\ref{eq:d1_F_torque}) and (\ref{eq:SW_Doppler}) show.

\section{Application}

 In this section, we apply the formula, Eq.~(\ref{eq:result}), to topological insulator surface states, 
a 2D magnetic Rashba system, and a magnetic Weyl semimetal,  
and calculate $\tilde {\bm D}_i^{(2)}$ and $\tilde {\bm D}_i$ analytically.  
 In the Rashba model, ${\bm M}$ in the coefficients ($\tilde {\bm D}_i^{(2)}$ and $\tilde {\bm D}_i$)
 is assumed perpendicular to the 2D plane. 
 While the general formula presented in the preceding section is valid at any temperature, 
we focus on absolute zero, $T=0$, in this section.  
 Applications to systems with weak SOC are described in Appendix \ref{weak_SOC}.

\subsection{General expression for two-band models}

 The models studied in this section are (essentially) described by $2 \times 2$ matrices 
in spin space, and the Hamiltonian has the form,    
\begin{align}
 H_M &= \sum_{\bm k} c_{\bm k}^{\dagger} 
  (\varepsilon_{\bm k} +  {\bm \lambda}_{\bm k} \cdot {\bm \sigma} + {\bm M} \cdot {\bm \sigma} )  \, 
   c_{\bm k}  , 
\label{eq:H_M_general}
\end{align}
where $c_{\bm k} = {}^t (c_{\bm{k}\uparrow}, c_{\bm{k} \downarrow})$ 
is the two-component electron operator, $\varepsilon_{\bm k}$ is the (scalar) energy, 
and ${\bm \lambda}_{\bm k}$ is the \lq\lq spin-orbit field''.  
 The magnetization $\bm{M}=(M_x,M_y,M_z)$ is assumed static and uniform, 
which is sufficient for the calculation of the coefficient $D_i^\alpha$ based on Eq.~(\ref{eq:result}).

 The $2 \times 2$  Green's function $G_M$ for $H_M$ is 
\begin{eqnarray}
 G_M ({\bm k},i\varepsilon_m) 
&=&  \frac{g_0 + {\bm g}_M \cdot {\bm \sigma} }{D_M}  , 
\label{eq:G_M_2x2_g}
\end{eqnarray}
where 
$g_0 = i\varepsilon_m + \mu - \varepsilon_{\bm k}$, 
${\bm g}_M = {\bm \lambda}_{\bm k} + {\bm M}$, 
and $D_M = g_0^2 - {\bm g}_M^2 $. 
 The formula (\ref{eq:result_D_tilde_2}) then reduces to 
\begin{equation}
  \tilde {\bm D}_i^{(2)} 
= T \sum_m \sum_{\bm k} \frac{(\partial_i g_0) {\bm g}_M - g_0 (\partial_i {\bm g}_M)}{D_M^2} .  
\label{eq:D_2x2}
\end{equation}

\subsection{Magnetic topological insulator} 

 As a first example, we consider 2D Dirac fermions on a surface of topological 
insulator,\cite{TI_review,TI_spin} which are described by the Hamiltonian, 
\begin{align}
 H_M &= \sum_{\bm k} c_{\bm k}^{\dagger}  
  \left\{ - v_{\rm F} (\hat{z} \times \bm{k}) \cdot {\bm \sigma} + {\bm M} \cdot {\bm \sigma} \right\}  
   c_{\bm k} . 
\label{eq:H_M_TI}
\end{align}
 In this model, the constant $M_x$ and $M_y$ are gauge degrees of freedom 
and can be eliminated by shifting $k_x$ and $k_y$. 
 Hence we set $M_x = M_y = 0$ from the first. 
 (Note that the spatial gradient in the DMI has already been extracted.)
 The electron spectrum is of the Dirac type, $\pm \sqrt{(v_{\rm F} k)^2 + M_z^2}$, 
with an \lq\lq exchange gap'' $2|M_z|$.

 Noting that Eq.~(\ref{eq:D_2x2}) reduces to 
\begin{eqnarray}
  \tilde D_i^{(2), \alpha} &=&  v_{\rm F}  \, \varepsilon_{i\alpha}  
         T\sum_m \sum_{\bm k} \frac{g_0}{D_M^2} , 
\end{eqnarray}
where $\varepsilon_{i \alpha} = \varepsilon_{i \alpha z}$, 
we take the Matsubara ($m$-) sum and perform the ${\bm k}$-integration as 
\begin{eqnarray}
 T\sum_m \sum_{\bm k} \frac{g_0}{D_M^2}  
&=&  -  \frac{{\rm sgn} \mu }{8\pi  v_{\rm F}^2} \, \Theta (|\mu| > |M_z|) . 
\end{eqnarray}
 This is finite in the doped state, $|\mu| > |M_z|$, but vanishes in the insulating state,  
$|\mu| < |M_z|$, as indicated by the Heaviside step function $\Theta$. 
 Thus, 
\begin{eqnarray}
   D_i^{(2), \alpha} 
&=& - \varepsilon_{i\alpha} \, \frac{{\rm sgn} \mu }{8 \pi  v_{\rm F}}  \times 
        \left\{ \begin{array}{cc} M^2 & (|\mu| > |M_z|)  \\ 
                 0 & (|\mu| < |M_z|)  \end{array} \right.   . 
\end{eqnarray}
 This agrees with the result in Ref.~\onlinecite{Wakatsuki2015} if $M^2$ is replaced 
by $J_\perp J_\parallel$.\cite{com1}  
 This DMI coefficient applies to the spin-wave dispersion. 
 The factor $\varepsilon_{i\alpha}$ indicates that a maximal Doppler shift is attained 
when the spin-wave propagation direction is perpendicular to the in-plane magnetization direction 
[see Eq.~(\ref{eq:SW_Doppler})].

\begin{figure}[t]
\includegraphics[width=70mm]{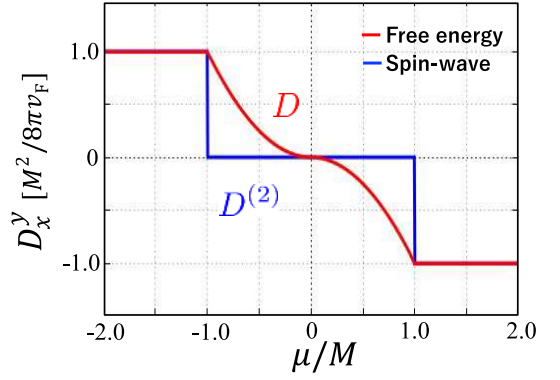}
\caption{The DMI coefficients, $\tilde D_x^y$  and $\tilde D_x^{(2), y}$, 
mediated by topological insulator surface states plotted against the chemical potential ($\mu$). 
}
\label{fig:DMI_TI_Rashba}
\end{figure}

 The DMI coefficient ${\bm D}_i$ in free energy is obtained by integrating 
$\tilde {\bm D}_i^{(2)}$ with respect to $M$, 
\begin{eqnarray}
 D_i^\alpha &=&  - \varepsilon_{i\alpha} \, \frac{{\rm sgn} \mu }{4\pi  v_{\rm F}}
        \int_0^M M' dM' \,  \Theta (|\mu| > |M_z'|)  . 
\end{eqnarray}
 By writing $M_z = M \cos\theta$, 
where $\theta$ is the angle that ${\bm M}$ makes with surface normal, 
the integration is carried out as 
\begin{align}
&   \int_0^M M' dM' \,  \Theta (|\mu| > M' |\cos \theta |) 
\nonumber \\
&=   \int_0^{{\rm min} \{ M , \frac{|\mu|}{|\cos\theta|} \}} M' dM' 
=   \frac{1}{2} {\rm min} \left\{ M^2 , \frac{\mu^2}{\cos^2\theta} \right\} , 
\end{align}
and we obtain 
\begin{eqnarray}
 D_i^\alpha 
&=& - \varepsilon_{i\alpha} \, \frac{{\rm sgn} \mu }{8\pi  v_{\rm F}}  \times 
        \left\{ \begin{array}{cc} M^2 & (|\mu| > |M_z|)   \\  
                \displaystyle \frac{\mu^2}{\cos^2\theta} & (|\mu| < |M_z|)  \end{array} \right.   . 
\label{eq:D_TI}
\end{eqnarray}
 Interestingly, this is finite even when $\mu$ lies in the gap ($|\mu| < |M_z|$). 
 This feature is different from $D_i^{(2), \alpha}$, which vanishes there. 
 Thus we have a first example that demonstrates ${\bm D}_i^{(2)} \ne {\bm D}_i$.

 Though $D_i^\alpha$ is typically nonzero even in the gap, it is an odd function of $\mu$ and vanishes 
at $\mu=0$ (i.e., in the ground or equilibrium state). 
 A finite DMI in the gap, realized at $\mu \ne 0$, may be probed if the chemical potential is tuned 
by, e.g., gating or impurity doping.

 The mathematical reason why ${\bm D}_i$ is nonzero (while ${\bm D}_i^{(2)} = {\bm 0}$) in the gap 
is that ${\bm D}_i$ at given $M$ is contributed from ${\bm D}_i^{(2)}$ 
at smaller $M$ because of the $M$-integration. 
 That is, even when $\mu$ lies in the gap at some given $M$, if it is nonzero ($\mu \ne 0$), 
it enters the band at smaller $M$, at which ${\bm D}_i^{(2)}$ is nonzero and contributes to ${\bm D}_i$. 
 Physically, it means that twisting the magnetization changes the electronic energy 
in a chirality-dependent way even in the insulating state.

 The factor $\varepsilon_{i\alpha}$ indicates that a N\'eel type spin twist is favored.

\subsection{Rashba ferromagnet} 
\label{Rashba}

\begin{figure*}[t]
\includegraphics[width=180mm]{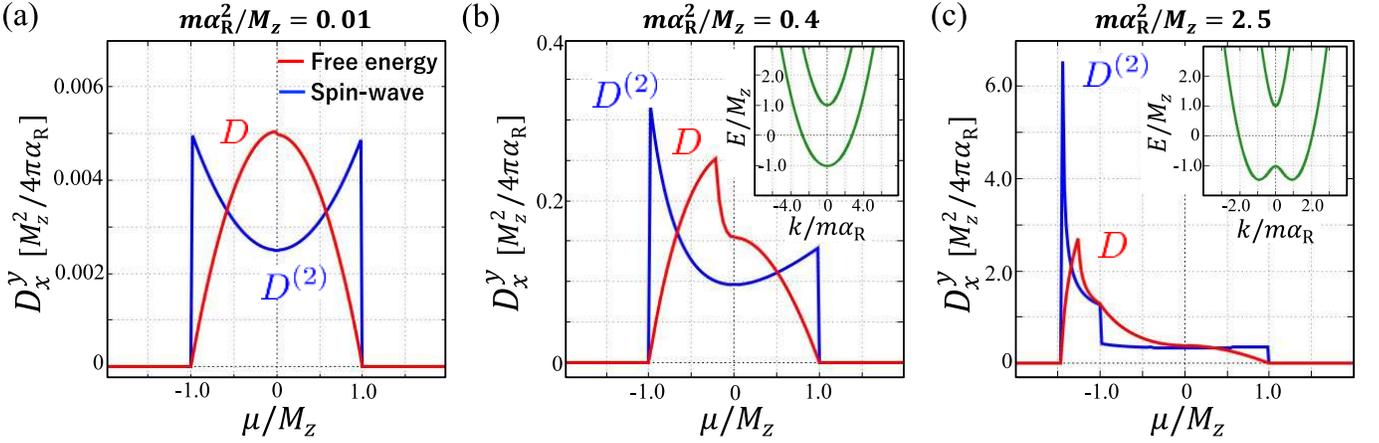}
\caption{The DMI coefficients, $D_x^y$ and $D_x^{(2), y}$, in a Rashba ferromagnet 
plotted as functions of the chemical potential $\mu$. The insets show the band structures. 
(a) At first order in $\alpha_{\mathrm{R}}$. 
($m \alpha_{\mathrm{R}}^2/M_z =0.01$ is used for the plot.) 
 Note that a slight discontinuity-like structure in $D_x^y$ at $\mu=0$ is not a numerical error; 
see Fig.~\ref{fig:DMI_spincurrent_Rashba} (a). 
(b) For a ``ferromagnetic'' band structure with $m \alpha_{\mathrm{R}}^2/M_z =0.4$. 
(c) For a wine-bottle type band structure with $m \alpha_{\mathrm{R}}^2/M_z = 2.5$. 
}
\label{fig:DMI_band_Rashba}
\end{figure*}

 As a next example, we consider 2D electrons with Rashba SOC, 
\begin{align}
 H_M &= \sum_{\bm k} c_{\bm k}^{\dagger}  
  \left\{ \varepsilon_{\bm k} - \alpha_{\rm R} (\bm{k} \times \bm{\sigma})_z   
   + {\bm M} \cdot {\bm \sigma} \right\}  
   c_{\bm k} , 
\label{eq:H_Rashba_M}
\end{align}
where $\varepsilon_{\bm k} = \hbar^2 {\bm k}^2/2m$ with electron (or effective) mass $m$, 
and $\alpha_{\rm R}$ is the Rashba constant. 
 In evaluating the DMI coefficients, we assume ${\bm M}=(0,0,M_z)$ is perpendicular to the 2D plane.  
 The electron energy is given by
\begin{align}
E_{\bm{k}\pm} = \varepsilon_{\bm k} \pm \sqrt{ (\alpha_{\rm R} k)^2 + M_z^2} .
\end{align}
 For $M_z=0$, there is a Dirac point at ${\bm k}={\bm 0}$, but a finite $M_z$ eliminates it 
by opening an exchange gap of magnitude $2 M_z$. 
 (In this subsection, we assume $M_z \geq 0$ for simplicity, 
 but the results do not depend on the sign of $M_z$.)  
 For $\alpha_{\mathrm{R}}< \sqrt{M_z/m}$, the band structure is similar to that of ferromagnets, 
which is dominated by the exchange splitting [see the inset of Fig.~\ref{fig:DMI_band_Rashba} (b)]. 
 For $\alpha_{\mathrm{R}}> \sqrt{M_z/m}$, 
the shape of the lower band becomes wine-bottle-like [Fig.~\ref{fig:DMI_band_Rashba} (c), inset].
 We call the former band structure ``ferromagnetic type'' and the latter ``wine-bottle type''.
 The bottom of the lower band is at energy $\varepsilon_{\rm min} =  - M_z$ for ferromagnetic type, 
and $\varepsilon_{\rm min} = - \frac{1}{2} \left( m \alpha_{\rm R}^2 + \frac{M_z^2}{m \alpha_{\rm R}^2} \right)$ 
for wine-bottle type.

 To study the DMI coefficients, it may be instructive to start with the lowest-order 
(i.e., first-order) contribution with respect to $\alpha_{\rm R}$. 
 In this case, the band structure is of ferromagnetic type, 
and the DMI coefficient is given by the spin current\cite{Kikuchi2016}  (see Appendix \ref{weak_SOC}). 
 Using Eq.~(\ref{eq:D_weak_soc_2}) 
with $\lambda_i^\alpha = - \alpha_\mathrm{R} \varepsilon_{i\alpha} $, $d=2$,  
and $\nu_{\uparrow}=\nu_{\downarrow} = m / 2\pi$ (density of states), 
the coefficient of DM free energy is obtained as 
\begin{align}
\label{DMI_energy_Rashba}
 D^{\alpha}_i
= \varepsilon_{i \alpha} \frac{\alpha_\mathrm{R} m M}{8\pi} 
\begin{cases}
1 - \left(\mu/M \right)^2, & \left| \mu \right| < M 
\\
0, & \left| \mu \right| > M   
\end{cases}  , 
\end{align}
where $M=|\bm{M}|$ is the magnitude of the magnetization. 
(The results in this paragraph holds for arbitrary ${\bm M}=(M_x,M_y,M_z)$.)
 This agrees with the result obtained by Ado et al.\cite{Ado2018}  
 It is finite only in the half-metallic state, in which the chemical potential crosses only the lower band. 
 The DMI in spin-wave dispersion is obtained as [Eq.~(\ref{eq:D^2_eq1})]
\begin{align}
D^{(2),\alpha}_i
= \varepsilon_{i \alpha} \frac{ \alpha_\mathrm{R} m M}{16\pi} 
\begin{cases}
 1  +\left (\mu/M \right)^2, & \left| \mu \right| < M 
\\
 0, & \left|\mu\right| > M 
\end{cases}  . 
\end{align}
 These results, also plotted in Fig.~\ref{fig:DMI_band_Rashba} (a), 
clearly demonstrate the distinction between $D^\alpha_i$ and $D^{(2),\alpha}_i$ (for $| \mu| < M$).

\begin{figure*}
\includegraphics[width=180mm]{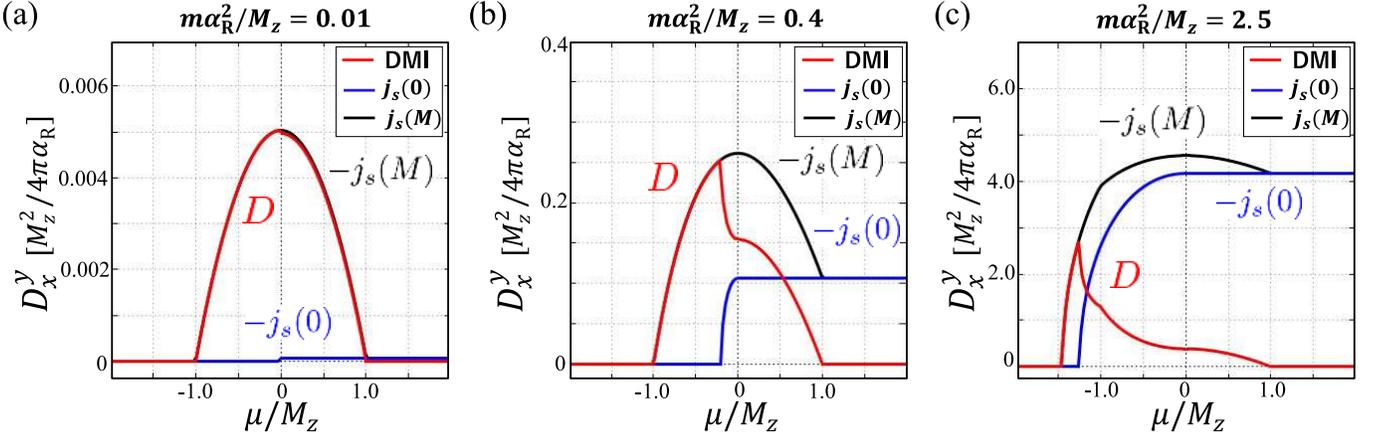}
\caption{Equilibrium spin current, $j_{\rm s}$ ($ = j_{{\rm s}, x}^y$), and the DMI coefficient calculated from 
$D_x^y = - \{ j_{\rm s} (M) - j_{\rm s} (0) \}$ [Eq.~(\ref{eq:D_spin_current})] 
for a Rashba ferromagnet. 
 The parameters are the same as in Fig.~\ref{fig:DMI_band_Rashba}. 
(a) $m \alpha_{\mathrm{R}}^2/M_z =0.01$. 
(b) $m \alpha_{\mathrm{R}}^2/M_z = 0.4$ (ferromagnetic type). 
(c) $m \alpha_{\mathrm{R}}^2/M_z =2.5$ (wine-bottle type). 
}
\label{fig:DMI_spincurrent_Rashba}
\end{figure*}

 To include higher-order effects with respect to $\alpha_{\rm R}$, we use Eq.~(\ref{eq:D_2x2}). 
 For a perpendicular magnetizatoin, the ${\bm k}$-integration can be performed analytically   
to obtain $D_i^{(2),\alpha} $ (for spin waves).  
 The result is 
\begin{widetext}
\begin{align}
 D_i^{(2),\alpha} (M_z) 
 &= \varepsilon_{i\alpha}  \frac{M_z^2}{4\pi \alpha_{\mathrm{R}}} \left[ 
   \frac{m  \alpha_{\mathrm{R}}^2-\varepsilon_{k_-}}
          {\left|  \sqrt{ (\alpha_{\mathrm{R}} k_{\mathrm{F}-})^2 + M_z^{2}} -m  \alpha_{\mathrm{R}}^2 \right|}
 - \frac{m  \alpha_{\mathrm{R}}^2-\varepsilon_{k_+}}
           {\left|\sqrt{ (\alpha_{\mathrm{R}} k_{\mathrm{F}+})^2 + M_z^{ 2}} +m  \alpha_{\mathrm{R}}^2  \right|} \right] 
\nonumber\\
&\quad 
 +\varepsilon_{i\alpha} \frac{M_z^2}{4\pi \alpha_{\mathrm{R}}} \left[ \frac{ \sqrt{ (\alpha_{\mathrm{R}} k_{\mathrm{F}-})^2 + M_z^{ 2}}-M_z}{ m \alpha_{\mathrm{R}}^2}-\frac{\sqrt{ (\alpha_{\mathrm{R}} k_{\mathrm{F}+})^2 + M_z^{2}}-M_z}{ m \alpha_{\mathrm{R}}^2} \right] , 
\label{eq:D2_Rashba}
\end{align}
\end{widetext}
for $\mu > \varepsilon_{\rm min}$, whereas it trivially vanishes for $\mu < \varepsilon_{\rm min}$. 
 Here, $\varepsilon_{k_\pm} = \hbar^2  k_{\mathrm{F}\pm}^2/2m $, 
and $k_{\mathrm{F}\pm}$ are real positive solutions of 
\begin{align}
  \frac{\hbar^2 k_{\mathrm{F}\pm}^2 }{2m} \pm \sqrt{ (\alpha_{\mathrm{R}} k_{\mathrm{F}\pm})^2 +M_z^2}  
= \mu ,
\end{align}
with $k_{{\rm F}+}$ ($k_{{\rm F}-}$) for the upper (lower) band. 
 Depending on the Fermi-surface morphology, Eq.~(\ref{eq:D2_Rashba}) should be interpreted as follows.

1,  For $\mu > M_z$, the contributions from the two Fermi surfaces (each from the upper and lower bands) 
cancel exactly, resulting in $D_i^{(2),\alpha}  = 0$. 

2. For $\mu < M_z$, there is no solution for $k_{{\rm F}+}$, and the terms that contain $k_{{\rm F}+}$ 
 in Eq.~(\ref{eq:D2_Rashba}) should be set to zero.  

3. For a wine-bottle type band, there is one more region, $\varepsilon_{\rm min} < \mu < -M_z$, 
where one finds two solutions for $k_{{\rm F}-}$. 
 In Eq.~(\ref{eq:D2_Rashba}), the terms that contain $k_{{\rm F}-}$ should then be understood as 
the sum of contributions from the two solutions (whereas the terms that contain $k_{{\rm F}+}$ are set to zero).

 The results are plotted in Fig.~\ref{fig:DMI_band_Rashba} (b) and (c) by blue lines.

 To obtain ${\bm D}_i$ (for free energy), we first performed the following integration numerically, 
\begin{align}
  {\bm D}_i (M_z) = \int_0^{M_z} 2 M'_z dM'_z   \tilde {\bm D}_i^{(2)} (M'_z) , 
\label{eq:D_Rashba}
\end{align} 
which required a careful division into several intervals    
(see the above result on $\tilde {\bm D}_i^{(2)}$).  
 The results are plotted in Fig.~\ref{fig:DMI_band_Rashba} (b) and (c) by red lines.
 We see that it also vanishes for $\mu > M_z$, where both bands have a Fermi surface. 
 This feature was reported in Ref.~\onlinecite{Ado2018} at the lowest-order in $\alpha_{\rm R}$. 
 Here, we have shown it to all orders in $\alpha_{\rm R}$.

\begin{figure*}[t]
\includegraphics[width=120mm]{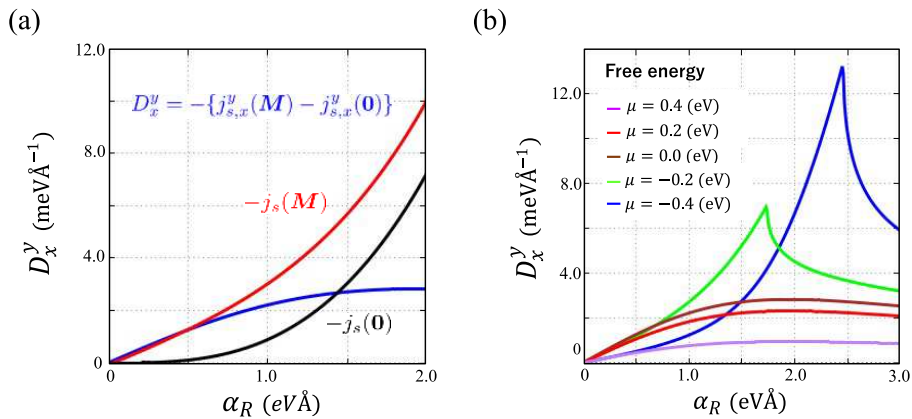}
\caption{The $\alpha_{\rm R}$ dependence of the DMI coefficient $D_x^y$ calculated from 
the spin-current formula, 
$D_x^y = - \{ j_{{\rm s},x}^y (M) - j_{{\rm s}, x}^y (0) \}$ [Eq.~(\ref{eq:D_spin_current})], 
for a Rashba ferromagnet.
 The parameters are chosen as $M_z = 0.5~\mathrm{eV}$, 
as in Ref.~\onlinecite{Freimuth2017}. 
 (a) For $\mu = 0$. Equilibrium spin currents are also shown. 
 This is compared to Fig.~2 of Ref.~\onlinecite{Freimuth2017}. 
 A slight difference is seen in $D_x^y$ from that in Ref.~\onlinecite{Freimuth2017},  
 which is likely to come from the different formula exploited. 
 (b) For several choices of $\mu$.
}
\label{fig:DMI_Rashba_alpha}
\end{figure*}

 Interestingly, for the present case of perpendicular magnetization, 
${\bm D}_i$ can be expressed by the equilibrium spin-current density, ${\bm j}_{{\rm s},i}$. 
 More precisely, their perpendicular components (perpendicular to ${\bm M}$) are mutually related as   
\begin{align}
  {\bm D}_i^\perp  &= - \{ {\bm j}_{{\rm s},i}^\perp (M) - {\bm j}_{{\rm s},i}^\perp (0) \} . 
\label{eq:D_spin_current}
\end{align}
 This is derived in Appendix \ref{spin_current_formula} as Eq.~(\ref{eq:app:D_spin_current}).   
 This enables us to have analytical expressions of ${\bm D}_i^\perp$, 
as displayed in Appendix \ref{Rashba_spin_current}. 
 Each term in Eq.~(\ref{eq:D_spin_current}) is plotted in Fig.~\ref{fig:DMI_spincurrent_Rashba}. 
 As seen by the red curves, the agreement between 
Fig.~\ref{fig:DMI_band_Rashba} (calculated by Eq.~(\ref{eq:D_Rashba})) 
and Fig.~\ref{fig:DMI_spincurrent_Rashba} (calculated by Eq.~(\ref{eq:D_spin_current})) is perfect, 
demonstrating that the spin-current picture works nicely. 
 The following features are seen: 

1. At small $\alpha_{\rm R}$ [Fig.~\ref{fig:DMI_spincurrent_Rashba} (a)], 
the spin current at $M=0$ is negligibly small, and the DMI fully reflects the equilibrium spin current at $M$. 

2. At larger $\alpha_{\rm R}$ [Fig.~\ref{fig:DMI_spincurrent_Rashba} (b), (c)], 
the spin current develops even at $M=0$ and affects the DMI coefficient. 
 In particular, the abrupt drop in $D_x^y$ with a cusp is due to the subtraction of the spin current at $M=0$. 

3. When both bands are occupied ($\mu > M_z$), the spin currents are nonvanishing but independent of $M$. 
  Therefore, the DMI coefficient vanishes.

 In Ref.~\onlinecite{Freimuth2017}, it was shown by explicit calculations that ${\bm D}_i$ deviates from 
$- {\bm j}_{{\rm s},i} (0)$ at large $\alpha_{\rm R}$ when the nonlinear dependence on $\alpha_{\rm R}$ 
becomes appreciable. 
 In Fig.~\ref{fig:DMI_Rashba_alpha} (a), we show the same plot but using our spin-current formula, 
Eq.~(\ref{eq:D_spin_current}), for ${\bm D}_i$. 
 Our plot for ${\bm D}_i$ (blue line) agrees well with the one presented 
in Ref.~\onlinecite{Freimuth2017} (not shown); 
in a closer look, there is a slight disagreement, 
which may be attributed to the use of different DMI formula here and in Ref.~\onlinecite{Freimuth2017}. 
 The DMI coefficient for several choices of $\mu$ are plotted in Fig.~\ref{fig:DMI_Rashba_alpha} (b).

\subsection{Magnetic Weyl semimetal}

 Finally, we consider a magnetic Weyl semimetal. 
 We consider a model with two Weyl cones of opposite chirality,\cite{Weyl_review, Kurebayashi2021} 
\begin{align}
 H_M &= \sum_{{\bm k},\tau} c_{{\bm k},\tau}^{\dagger} (\tau v_{\rm F} {\bm k} \cdot {\bm \sigma} 
      + {\bm M} \cdot {\bm \sigma} - \tau  \Delta ) 
   c_{{\bm k},\tau} , 
\label{eq:H_M_Weyl}
\end{align}
where $\tau = \pm 1$ specifies the chirality of the Weyl cones. 
 The magnetization ${\bm M}$ and the chiral asymmetry parameter $\Delta$ 
determine the relative shift of the two Weyl cones 
in the momentum and energy directions, respectively. 
 The latter ($\Delta$) arises when the system lacks spatial inversion symmetry. 
 The model in the present form was used recently in a microscopic study of various types of 
(current- and charge-induced) spin torques.\cite{Kurebayashi2021}

 Using the formula (\ref{eq:D_2x2}) with ${\bm g}_M = \tau v_{\rm F} {\bm k} + {\bm M}$, 
we perform the ${\bm k}$-integral by shifting the integration variable ${\bm k}$ 
to eliminate ${\bm M}$, such that ${\bm g}_M \to \tau v_{\rm F} {\bm k}$. 
 This shift is done in a chirality-dependent way,\cite{com2}  
which may be justified if the two Weyl cones (with opposite chiralities) can be treated independently, 
e.g., if the scattering between the two Weyl cones can be neglected.  
 As a result, the DMI coefficient, calculated from Eq.~(\ref{eq:D_2x2}), 
\begin{align}
 \tilde D_i^{(2), \alpha} 
&= \sum_{\tau = \pm 1} \frac{\tau (\mu + \tau \Delta)}{8\pi^2 v_{\rm F}^2 } \delta_{i\alpha}   
= \frac{\Delta}{4\pi^2 v_{\rm F}^2 } \delta_{i\alpha}  , 
\label{eq:result_Weyl} 
\end{align}
is independent of ${\bm M}$. 
 Therefore, $\tilde {\bm D}_i$ and $\tilde {\bm D}_i^{(2)}$ coincide, 
$\tilde {\bm D}_i = \tilde {\bm D}_i^{(2)}$; see Eq.~(\ref{eq:D^2_eq1}). 
 This feature is consistent with ${\rm div}_M \tilde {\bm D}_i = 0$ (see Appendix \ref{Remarks}).  
 The contributions from opposite-chirality branches tend to cancel each other 
[see the middle in Eq.~(\ref{eq:result_Weyl})], 
and the nonvanishing DMI is obtained only when $\Delta \ne 0$, 
consistent with the symmetry requirement. 
 We also note that it is independent of $\mu$. 
 The factor $\delta_{i\alpha}$ indicates that the spin-wave Doppler shift is maximal  
when the propagation direction is parallel (or antiparallel) to the magnetization. 
 The DM free energy is given by 
\begin{eqnarray}
 F_{\rm DM} 
&=&  - \frac{\Delta}{4\pi^2 v_{\rm F}^2 }  \int d{\bm r}  \,  {\bm M} \cdot ({\bm \nabla} \times {\bm M}) , 
\label{eq:DMI_Weyl}
\end{eqnarray}
which favors a Bloch-type spin twist.

 Here, we point out that the DMI in the Weyl system arises as a chiral anomaly. 
 To see this, it is convenient to rephrase the terminology with the language of Weyl fermions:  
(i) The electron spin ${\bm \sigma}$ is a chiral (or axial-vector) current, 
${\bm j}^5 = v_{\rm F} {\bm \sigma}$. 
(ii) The magnetization ${\bm M}$ acts as a chiral vector potential, 
${\bm A}^5 = - v_{\rm F}^{-1} {\bm M}$, 
and $\Delta$ acts as a chiral scalar potential, $A_0^5 = \Delta$ 
(with metric $-+++$).  
 See Ref.~\onlinecite{HK2021} for a quick overview.

\begin{figure}[t]
\includegraphics[width=90mm]{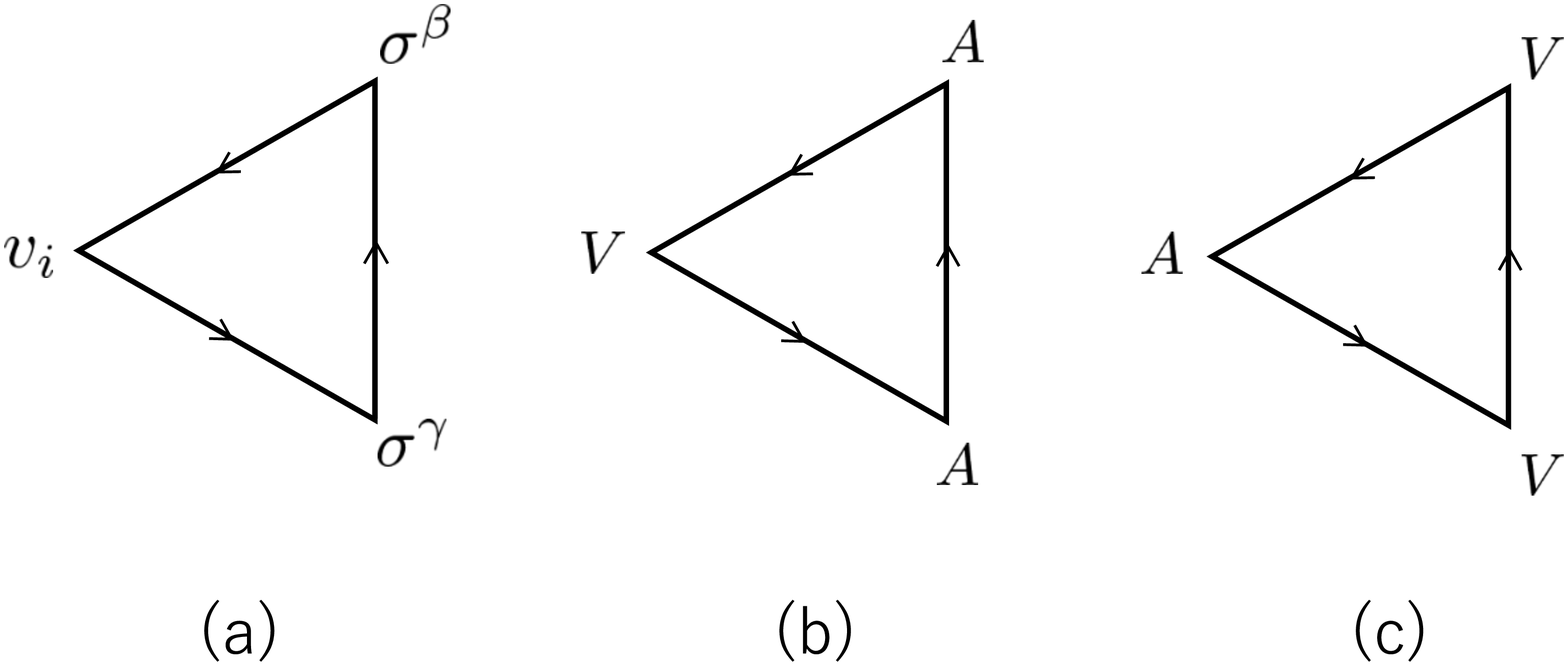}
\vskip -13mm
 \caption{
 Feynman diagrams.  
 (a) DMI coefficient for general cases. (b) DMI coefficient for Weyl semimetal.  
 (c) Chiral anomaly for Weyl semimetal. 
 In (b) and (c), $V$ represents velocity (vector current) and $A$ represents spin (axial current). 
}
\label{fig:triangle}
\end{figure}

 With these terminologies, the DMI is expressed by a triangle diagram with $VAA$ vertices 
[Fig.~\ref{fig:triangle} (b)], 
where $V$ means the (vector) current vertex ${\bm j} = \tau v_{\rm F} {\bm \sigma}$, 
and $A$ means the axial(-vector) current vertex ${\bm j}^5 = v_{\rm F} {\bm \sigma}$. 
 On the other hand, the chiral anomaly is expressed by a triangle diagram with $VVA$ vertices 
[Fig.~\ref{fig:triangle} (c)]. 
 The former ($VAA$) diagram usually vanishes. 
 However, in the presence of explicit chiral-symmetry breaking ($\Delta \ne 0$), 
the $VAA$ process mixes into the $VVA$ process and acquires a finite contribution.

 Let us look at the chiral anomaly term explicitly.\cite{Zyuzin2012,Liu2013} 
 Following Ref.~\onlinecite{Zyuzin2012}, one can derive an effective action 
(i.e., time integral of effective Lagrangian) using the Fujikawa method.\cite{Fujikawa_book}  
 Here, however, we are faced with the case in which ${\bm M}$ varies spatially. 
 To study such a situation, we divide ${\bm M}$ into the average part ($\bar {\bm M}$)  
and the fluctuation part ($\delta {\bm M}$), ${\bm M} = \bar {\bm M} + \delta {\bm M}$, 
where $\bar {\bm M}$ is constant and does not depend on space and time. 
 The same is assumed for $\Delta = \bar \Delta + \delta \Delta$.

 Eliminating the averages ($\bar {\bm M}$ and $\bar \Delta$) by a chiral gauge transformation, 
as done in Ref.~\onlinecite{Zyuzin2012}, 
and evaluating the associated change of path integral measure, we obtain the real-time action as 
\begin{align}
 S_\theta &= \frac{1}{32 \pi^2} \varepsilon^{\mu\nu\rho\lambda}  
   \int dt \int d{\bm r} \, \theta ({\bm r},t) F^5_{\mu\nu} F^5_{\rho\lambda} 
\nonumber \\
&= - \frac{1}{8 \pi^2} \varepsilon^{\mu\nu\rho\lambda}  
   \int dt \int d{\bm r} \, (\partial_\mu \theta ) A^5_\nu  \partial_\rho A^5_\lambda  . 
\label{eq:S_theta_1}
\end{align}
 Here, $F^5_{\mu\nu} = \partial_\mu A_\nu^5 - \partial_\nu A_\mu^5$ 
is the chiral electromagnetic field, given explicitly as  
$F^5_{ij} = - v_{\rm F}^{-1} \varepsilon_{ijk} ({\bm \nabla} \times {\bm M})^k$ and 
$F^5_{0i} = - \partial_i \Delta - v_{\rm F}^{-1} \dot M_i $. 
 (The familiar chiral anomaly term is obtained if $F^5_{\mu\nu} F^5_{\rho\lambda}$ is replaced 
by real electromagnetic fields, $F_{\mu\nu} F_{\rho\lambda}$.) 
 The \lq\lq axion field'' $\theta$ is determined by the averaged quantities,
$\theta ({\bm r},t) = 2( v_{\rm F}^{-1} \bar {\bm M} \cdot {\bm r} - \bar \Delta t)$. 
 When ${\bm M}$ is static and $\Delta$ is uniform, Eq.~(\ref{eq:S_theta_1}) reduces to 
\begin{align}
 S_\theta 
&=  \frac{\bar\Delta}{4\pi^2 v_{\rm F}^2} \int dt \int d{\bm r} \, 
     {\bm M} \cdot ({\bm \nabla} \times {\bm M}) . 
\label{eq:S_theta_2}
\end{align}
 From this, an effective free energy can be obtained by removing the time integral 
and attaching a minus sign; the result agrees with Eq.~(\ref{eq:DMI_Weyl}). 
 Therefore, the DMI in magnetic Weyl semimetals with broken inversion symmetry 
can be considered to originate from the chiral anomaly.

 It may be worth noting that our DMI formula gives the same result, 
without recourse to such subtle manipulations as represented by the chiral anomaly. 
 This may not be so surprising in view of the fact that the chiral anomaly 
is well captured within the framework of perturbation theory.\cite{Adler1969}

\renewcommand{\arraystretch}{1.8}
\begin{table*}[t]
\begin{center}
\caption{The DMI coefficients ${\bm D}_i$ and ${\bm D}_i^{(2)}$ in ferromagnets 
with various types of SOC. 
 The coefficients $C_1$ and $C_2$ in ${\bm D}_i$ are, in general, 
functions of ${\bm M} = (M_x,M_y,M_z)$ or $(M, {\bm n})$, and  we defined $\tilde C_1 \equiv C_1 / M^2$. 
 The term with $C_2$ is parallel to ${\bm M}$, hence does not contribute to the DM free energy, 
but is necessary for the condition ${\rm div} \tilde {\bm D}_i = 0$ to be satisfied. 
} 
\begin{tabular}{|c||c|c|c|c|} \hline
  Model  &  \  ${\bm D}_i = (D_i^\alpha)$ \   
& \  Condition of ${\rm div} \tilde {\bm D}_i = 0$ \  
& \ ${\bm D}_i^{(2)} = (D_i^{(2), \alpha})$ \ 
& \ ${\bm n} \cdot {\bm D}_i^{(2)}$ \   
\\ \hline\hline 
\ Weyl \footnote{See Eq.~(\ref{eq:H_M_Weyl}) for the Hamiltonian, and Eq.~(\ref{eq:result_Weyl}) for 
   $C_1 = \frac{\Delta M^2}{4\pi^2 v_{\rm F}^2 }$. Here, $i, \alpha = x,y,z$.}  
& $C_1 \, \delta_{i\alpha}$ 
& $\frac{\partial \tilde C_1}{\partial M^\alpha} = 0$  \  $(\alpha = x,y,z)$ \  
& $C_1 \, \delta_{i\alpha}$  
& $C_1 \, \delta_{i\alpha} n^\alpha$   
\\ \hline 
\ MTI surface \footnote{See Eq.~(\ref{eq:H_M_TI}) for the Hamiltonian, and Eq.~(\ref{eq:D_TI}) for 
   $C_1 = - \, \frac{{\rm sgn} \mu }{8\pi  v_{\rm F}}  \frac{M^2}{M_z^2} {\rm min} \{ \mu^2, M_z^2\}$. 
    Here, $i, \alpha = x,y$.}   
&  $C_1 \, \varepsilon_{i\alpha}$ 
& \ $\frac{\partial \tilde C_1}{\partial M^\beta} = 0$ \  $(\beta = x,y)$ \  
& $ \frac{M}{2} \frac{\partial C_1}{\partial M} \varepsilon_{i\alpha} $  
& \ $\frac{M}{2} \frac{\partial C_1}{\partial M} \varepsilon_{i\beta} n^\beta $ \   
\\ \hline 
\ Rashba \footnote{See Eq.~(\ref{eq:H_Rashba_M}) for the Hamiltonian, Eq.~(\ref{eq:D_Rashba}) for $C_1$, 
and Eq.~(\ref{eq:D2_Rashba}) for $D_i^{(2), \alpha}$ with ${\bm M} \parallel \hat z$. 
   Here, $\alpha = x,y,z$, and $i, \beta = x,y$.}   
& \ $\varepsilon_{i\beta}(C_1 \delta_{\alpha\beta} + C_2 n^\alpha n^\beta)$  \  
& $ \frac{\partial \tilde C_1}{\partial M^\beta} = - \frac{n^\beta}{M^2} \frac{\partial C_2}{\partial M}$ 
& \  $\varepsilon_{i\beta}  \frac{M}{2} \left( \frac{\partial C_1}{\partial M} \delta_{\alpha\beta} 
   - M^2 n^\alpha \frac{\partial \tilde C_1}{\partial M^\beta} \right) $  \  
& \ \  $\varepsilon_{i\beta} \left( C_1 n^\beta -  \frac{1}{2} \frac{\partial C_1}{\partial n^\beta} \right)$ \ \   
\\ \hline 
\ Weak SOC 
  \footnote{See Eq.~(\ref{eq:H_weak_soc}) for the Hamiltonian, 
                and Eqs~(\ref{eq:D_weak_soc_1})-(\ref{eq:D_weak_soc_3}) for $C_1$. 
    We define ${\bm \lambda}_i^\parallel \equiv {\bm n} ({\bm n} \cdot {\bm \lambda}_i)$, and 
    ${\bm \lambda}_i^\perp \equiv {\bm \lambda}_i - {\bm \lambda}_i^\parallel$. }  
& \ $C_1 {\bm \lambda}_i^\perp + C_2 {\bm \lambda}_i^\parallel $ \ 
& \ $C_1 = \frac{M}{2} \frac{\partial C_2}{\partial M}$, \ 
     $\frac{\partial C_1}{\partial {\bm n}} = {\bm 0}$ \ 
& \ $C_1 {\bm \lambda}_i^\parallel + \frac{M}{2} \frac{\partial C_1}{\partial M}  {\bm \lambda}_i^\perp$ \   
& $C_1 ({\bm n} \cdot {\bm \lambda}_i^\parallel )$    
\\ \hline 
\end{tabular}
\label{table_I}
\end{center}
\end{table*}

\section{Discussion} 

\subsection{${\bm D}_i$ vs. ${\bm D}_i^{(2)}$} 

 We have emphasized the distinction between ${\bm D}_i$ and ${\bm D}_i^{(2)}$. 
 However, their difference may not be significant if ${\bm D}_i/M^2$ and ${\bm D}_i^{(2)}/M^2$ 
are weakly dependent on $M$. 
(If ${\bm D}_i$ and/or ${\bm D}_i^{(2)}$ are strictly proportional to $M^2$, 
as in the case of magnetic Weyl semimetal, one has ${\bm D}_i = {\bm D}_i ^{(2)}$.)
 In the experiment in Ref.~\onlinecite{Zhou2020}, the authors extracted the scaling 
$D_i^\alpha \propto M_{\rm s}^p$ ($M_{\rm s}$ : saturation magnetization) 
using the temperature as an implicit parameter.
 They obtained the exponent $p = 1.84 \pm 0.16$, which is consistent with, or not far from, $p=2$. 
 To see the difference between ${\bm D}_i$ and ${\bm D}_i^{(2)}$ experimentally, 
materials having $p$ significantly different from 2 are good candidates.

 Explicit relations between ${\bm D}_i$ and ${\bm D}_i^{(2)}$ that follow 
from ${\rm div} \tilde {\bm D}_i = 0$ are summarized in Table \ref{table_I} for several types of SOC. 
 (See Appendix \ref{Remarks} for the discussion.) 
 While ${\bm D}_i$ is expressed by a single quantity $C_1$ 
for a Weyl semimetal and magnetic topological insulator, two quantities ($C_1$ and $C_2$) are required 
for Rashba ferromagnet and for systems with general but weak SOC.  
 In all cases studied here, ${\bm D}_i^{(2)}$ can be expressed solely by $C_1$ 
(because $C_2$ is related to $C_1$ via ${\rm div} \tilde {\bm D}_i = 0$), 
hence $C_2$ is practically unnecessary. 
 This is reasonable since one can always take the DM free energy as a starting point, 
where only the $C_1$ term is relevant.

\subsection{Parallel vs. perpendicular components}

 In comparing ${\bm D}_i$ and ${\bm D}_i^{(2)}$, one needs to pay attention to 
their vectorial direction relative to ${\bm M}$, namely, to the components 
parallel or perpendicular to ${\bm M}$. 
 (We call the former parallel component and the latter perpendicular component.) 
  While the derived formula for ${\bm D}_i$ [Eq.~(\ref{eq:result})] and ${\bm D}_i^{(2)}$ [Eq.~(\ref{eq:result_D_tilde_2})] 
generally contain both parallel and perpendicular components, 
only the perpendicular (parallel) component is physically relevant for ${\bm D}_i$ (${\bm D}_i^{(2)}$), 
as we have seen in Sec.~\ref{phys}. 
 Let us denote them ${\bm D}_i^\perp$ and ${\bm D}_i^{(2),\parallel}$, respectively.

 As seen from Table I, ${\bm D}_i^\perp$ and ${\bm D}_i^{(2),\parallel}$ have the same coefficient 
for the magnetic Weyl semimetal and materials with weak SOC. 
 On the other hand, they have different coefficients for MTI surface electrons and Rashba ferromagnets. 
 In the latter cases, the DMI coefficient determined from the length scale of 
spin textures and that determined from spin-wave properties can be different.

 In practical calculations, such projections (to parallel or perpendicular components) can be done separately for ${\bm D}_i$ and ${\bm D}_i^{(2)}$, as in Eqs.~(\ref{eq:D_weak}) and (\ref{eq:D2_weak}). 
 On the other hand, if one is to use 
Eq.~(\ref{eq:DMI_relation_1}) to obtain ${\bm D}_i^{(2)}$ from ${\bm D}_i$, 
or Eq.~(\ref{eq:DMI_relation_2}) to obtain ${\bm D}_i$ from ${\bm D}_i^{(2)}$, 
both components need to be retained.

\subsection{Relation to equilibrium spin current} 

 Kikuchi et al.~pointed out that the DMI coefficient is given by the equilibrium spin-current density.\cite{Kikuchi2016} 
 Their \lq\lq spin-current formula'' holds at the first order in SOC, 
thus it is appropriate to systems with weak SOC. 
 Also, they assumed a parabolic electron dispersion as well as SOC that is linear in ${\bm k}$.

 In our formulation, such spin-current formula can be obtained if first-order terms in SOC 
are retained. 
 This is demonstrated in Appendix \ref{spin-current}, where one can see that the spin-current formula,  
\begin{align}
 {\bm D}_i^\perp (M)  = -   {\bm j}_{{\rm s},i} (M)  , 
\label{eq:spin_current_formula_1}
\end{align}
holds for arbitrary forms of electron dispersion and SOC. 
 Note that this is restricted to the perpendicular component of ${\bm D}_i$. 
 (The parallel component of the spin current vanishes,  
hence ${\bm j}_{{\rm s},i}^\perp = {\bm j}_{{\rm s},i}$, at the first order in SOC.)

 The equality between ${\bm D}_i^\perp$ and the equilibrium spin current can be extended 
to arbitrary strength of SOC (not just at the lowest order, but to all orders) 
for a Rashba ferromagnet with perpendicular magnetization.  
 In this case, the above relation is generalized to 
\begin{align}
 {\bm D}_i^\perp (M)  = - \left\{  {\bm j}_{{\rm s},i}^\perp (M) -  {\bm j}_{{\rm s},i}^\perp (0)  \right\} , 
\label{eq:spin_current_formula_2}
\end{align}
in which the spin current at ${\bm M} = {\bm 0}$ is subtracted. 
 Therefore, the DMI coefficient is equal to the {\it change} of the equilibrium spin current induced 
by the development of ${\bm M}$, rather than the \lq\lq preexisting'' spin current at ${\bm M} = {\bm 0}$. 
 It  is known that the equilibrium spin current at ${\bm M} = {\bm 0}$ is generally nonzero 
but starts at third order in SOC.\cite{Rashba2003,Tokatly2008,Droghetti2022} 
 Thus, Eq.~(\ref{eq:spin_current_formula_2}) reduces to Eq.~(\ref{eq:spin_current_formula_1}) 
when only the lowest- (i.e., first-)  order terms in SOC are considered. 
 Finally, we note that Eq.~(\ref{eq:spin_current_formula_2}) holds more generally (beyond the Rashba model) 
if the magnetization is perpendicular to the spin-orbit field. 
 A proof of this is given in Appendix \ref{spin_current_formula}.

\section{Summary}

 In this paper, we have developed a microscopic theory of the DMI mediated by band electrons. 
 In contrast to Moriya's original theory, which considered Mott insulators, 
the present one applies to metals, semiconductors, semimetals, and even to band insulators; 
in fact we have applied it to topological insulator surface states in its insulating state.

 We first derived a general formula for the coefficient of the DM free energy in terms of Green's functions. 
 We then pointed out that the DMI coefficient generally depends on physical quantities, 
i.e., whether we are looking at (equilibrium) spin structure, torque, or spin waves. 
 This distinction is important in systems with strong SOC, and spectacular examples are given by 
the Rashba ferromagnet in the half-metallic state (with one partially-occupied band and one empty band) 
and the MTI surface states in the insulating state. 
 Experimentally, it is expected that the difference between ${\bm D}_i$ and ${\bm D}_i^{(2)}$ 
can be significant when they deviate from the $\propto M^2$ behavior. 
 On the other hand, an exact  $\propto M^2$ behavior, hence the equality 
${\bm D}_i = {\bm D}_i^{(2)}$, has been found in magnetic Weyl semimetal. 
 This is probably an exceptional case among systems with strong SOC, 
as one might discuss through the analysis of the chiral anomaly.

 Also, we have generalized the spin-current formula for the coefficient of the DM free energy, 
which was known to hold in systems with weak SOC, to the case of strong SOC. 
 Under the assumption that the magnetization is perpendicular to the spin-orbit field, 
the DMI coefficient is proportional to the equilibrium spin current induced by 
the coupling to the magnetization.

\section*{acknowledgement}

 We would like to thank J.~J.~Nakane for his helpful discussion including his suggestion of Eq.~(\ref{eq:ab50}). 
 H.~K. is indebted to T. Ikeda for an early-stage collaboration, 
and to K.-J. Lee and C.-Y. You for stimulating discussion.   
 Y.~H. would like to take this opportunity to thank the \lq\lq Nagoya University Interdisciplinary Frontier Fellowship'' supported by Nagoya University and JST, the establishment of university fellowships towards the creation of science technology innovation, Grant Number JPMJFS2120.
 This work is also supported by JSPS KAKENHI Grant Numbers JP15H05702, JP17H02929, JP19K03744, 
and 21H01799.

\appendix

\begin{figure}[b]
\hspace*{15mm}
\includegraphics[width=70mm]{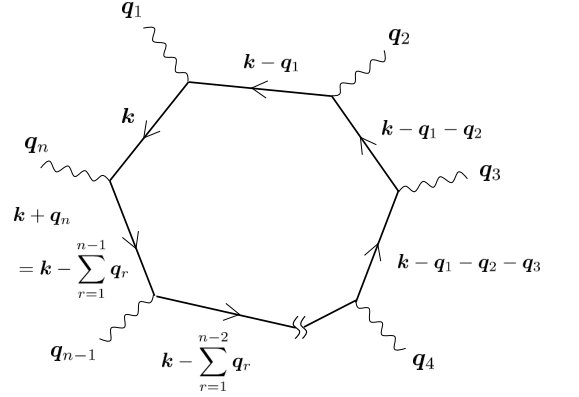}
 \caption{
 Feynman diagram at $n$th order with respect to the \lq\lq s-d'' exchange interaction, Eq.~(\ref{eq:app:eHsd}). 
 The wavy line with ${\bm q}_i$ represents ${\bm M}_{{\bm q}_i}$. 
 Note the momentum $-{\bm q}_r$ runs from the $n$th vertex to the $r$th vertex 
(counted clockwise from the ${\bm q}_1$ vertex). 
 For example, $-{\bm q}_3$ runs through the lower half of this diagram. 
}
\label{fig:Fig_diagram}
\end{figure}

\section{Derivation of general formula} 
\label{derivation}

\subsection{Basic derivation} 
\label{basic}

 In this Appendix, we derive the general expression for the DMI coefficient 
using the model described in Sec.~\ref{model}. 
 We write the exchange coupling as 
\begin{align} 
   H' =  \sum_{{\bm k}, {\bm q}} c_{\bm k}^\dagger {\cal H}'_{\bm q} c_{{\bm k}-{\bm q}}, 
\ \ \ \ \ 
   {\cal H}'_{\bm q} = {\bm M}_{\bm q} \cdot  {\bm \sigma}  ,
\label{eq:app:eHsd}
\end{align} 
and treat it as a perturbation but to an infinite order. 
 With the expansion formula, Eq.~(\ref{eq:F_start}), the free energy is calculated as 
\begin{widetext}
\begin{align}
 F &=  \frac{1}{\beta} \sum_{n=1}^\infty \frac{(-)^{n-1}}{n!} 
         \sum_{{\bm k}_1, \cdots, {\bm k}_n}  \sum_{{\bm q}_1, \cdots, {\bm q}_n}
         \int_0^\beta d\tau_1  \cdots \int_0^\beta d\tau_n \, 
     \langle {\rm T}_\tau [ 
           (c_{{\bm k}_1}^\dagger  {\cal H}'_{{\bm q}_1} \, c_{{\bm k}_1-{\bm q}_1})_{\tau_1} 
               \cdots
           (c_{{\bm k}_n}^\dagger  {\cal H}'_{{\bm q}_n} \, c_{{\bm k}_n-{\bm q}_n})_{\tau_n}  ]
      \rangle_{\rm conn}  
\nonumber \\
&= \frac{1}{\beta} \sum_{n=1}^\infty \frac{1}{n} 
         \sum_{{\bm k}_1, \cdots, {\bm k}_n}  \sum_{{\bm q}_1, \cdots, {\bm q}_n} 
         \int_0^\beta d\tau_1  \cdots \int_0^\beta d\tau_n  
\nonumber \\
&  \qquad   \times 
      {\rm tr} \bigl[\,  
        {\cal H}'_{{\bm q}_1} G_0 ({\bm k}_1-{\bm q}_1; \tau_1-\tau_2) \, \delta_{{\bm k}_1-{\bm q}_1,{\bm k}_2} 
        {\cal H}'_{{\bm q}_2} G_0 ({\bm k}_2-{\bm q}_2; \tau_2-\tau_3) \, \delta_{{\bm k}_2-{\bm q}_2,{\bm k}_3} 
         \cdots
        {\cal H}'_{{\bm q}_n} G_0 ({\bm k}_n-{\bm q}_n; \tau_n-\tau_1) \, \delta_{{\bm k}_n-{\bm q}_n,{\bm k}_1} 
\,\bigr] 
\nonumber \\
&= \sum_{n=1}^\infty \frac{1}{n} 
          \sum_{{\bm q}_1, \cdots, {\bm q}_{n-1}} 
      T \sum_m  \sum_{\bm k} 
      {\rm tr} \bigl[\,  
        {\cal H}'_{{\bm q}_1} G_0 ({\bm k} - {\bm q}_1, i\varepsilon_m) \, 
        {\cal H}'_{{\bm q}_2} G_0 ({\bm k} - {\bm q}_1 - {\bm q}_2, i\varepsilon_m)  
         \cdots
        {\cal H}'_{-({\bm q}_1+{\bm q}_2+ \cdots + {\bm q}_{n-1})} G_0 ({\bm k}, i\varepsilon_m)  \,\bigr]  ,  
\end{align}
\end{widetext}
where $G_0$ is given by Eq.~(\ref{eq:G_0}). 
 Note that there remains a symmetry factor of $1/n$ in the $n$th-order diagram, 
which is characteristic to free energy.\cite{AGD}  
 The external wave vectors ${\bm q}_1, {\bm q}_2, \cdots , {\bm q}_n$ are supplied 
from the magnetization, 
${\bm M}_{{\bm q}_1}, {\bm M}_{{\bm q}_2}, \cdots , {\bm M}_{{\bm q}_n}$. 
 We are interested in the terms which are first order in either of 
${\bm q}_1, {\bm q}_2, \cdots , {\bm q}_n$, 
and extract them from the Green's functions to first order. 
 For example, the $s$th Green's function (counted from left) having momentum  
${\bm k} - ({\bm q}_1 + {\bm q}_2 + \cdots + {\bm q}_s) $ is expanded as  
\begin{align}
& G_0 \left( {\bm k} - {\bm q}_1 - {\bm q}_2 - \cdots - {\bm q}_s , i\varepsilon_m \right) 
\nonumber \\
&\simeq  G_0 ({\bm k} , i\varepsilon_m) 
  - (q_{1,i} + q_{2,i}  + \cdots + q_{s,i}) \, \partial_i G_0 ({\bm k}, i\varepsilon_m) , 
\end{align}
up to $O(q)$. 
 It is convenient to first focus on a particular momentum ${\bm q}_r$ ($r = 1,2, \cdots, n-1$),  
collect all terms linear in ${\bm q}_r$, and then sum over $r$. 
 In $n$th-order diagrams, the momentum $-{\bm q}_r$ runs 
from the $r$th ${\cal H}'$ to the $n$th ${\cal H}'$, and its contributions are extracted as 
\begin{widetext}
\begin{align}
   \sum_{n=2}^\infty \frac{1}{n} 
         \sum_{{\bm q}_r} 
      T \sum_l  \sum_{\bm k}   (- q_{r,i}) \, 
      {\rm tr} \biggl[\,  
        ({\cal H}'_{\bm 0} G_0)^{r-1}  
         {\cal H}'_{{\bm q}_r}  \frac{\partial}{\partial k_i}  
         \left[ G_0  ({\cal H}'_{\bm 0}  G_0)^{n-r-1} \right] 
        ({\cal H}'_{-{\bm q}_r} G_0)  \,\biggr]  , 
\end{align}
\end{widetext}
where $G_0 = G_0 ({\bm k} , i\varepsilon_m)$. 
 Here, having extracted ${\bm q}_r$, we have set 
\begin{align}
  {\bm q}_i = {\bm 0}   \ \  {\rm for}  \ \  i \ne r, n , 
\label{eq:App:uniform_M}
\end{align}
(note the constraint, ${\bm q}_1 + {\bm q}_2 + \cdots + {\bm q}_n = {\bm 0}$)  
and wrote ${\cal H}'_{\bm 0}$ for ${\cal H}'_{{\bm q}={\bm 0}}$. 
 [The procedure (\ref{eq:App:uniform_M}) is actually not necessary, and will be relaxed later.] 
 Thus, ${\cal H}'_{\bm 0} = {\bm M} \cdot  {\bm \sigma}$, 
where ${\bm M} \equiv {\bm M}_{{\bm q}={\bm 0}}$ is the uniform magnetization. 
 By summing over $r$ and writing ${\bm q}_r$ as ${\bm q}$, 
we obtain 
\begin{widetext}
\begin{align}
 F &=  \sum_{n=2}^\infty \frac{1}{n} 
        \sum_{\bm q}  (- q_i) M_{\bm q}^\alpha  M_{-{\bm q}}^\beta  
      T \sum_l  \sum_{\bm k}    \sum_{r=1}^{n-1}  
       {\rm tr} \biggl[\,  
        ({\cal H}'_{\bm 0} G_0)^{r-1}  
         \sigma^\alpha  \frac{\partial}{\partial k_i}  
         \left[ G_0 ({\cal H}'_{\bm 0}  G_0)^{n-r-1} \right] 
        (\sigma^\beta G_0)  \,\biggr]   . 
\label{eq:App_5}
\end{align}
\end{widetext}
 To calculate the $r$-sum, we use the formula, 
\begin{align}
 \sum_{r=0}^N a^r b^{N-r} 
&= \frac{1}{N!} \left( \frac{d}{d\zeta}\right)^N 
      \frac{1}{1-\zeta a} \frac{1}{1-\zeta b} \Biggr|_{\zeta=0} , 
\label{eq:ab50}
\end{align}
with $N = n-2$. 
 The result is 
\begin{align}
& \sum_{r=1}^{n-1}  
       {\rm tr} \biggl[\,  
        ({\cal H}'_{\bm 0} G_0)^{r-1}  
         \sigma^\alpha  \frac{\partial}{\partial k_i}  
         \left[ G_0 ({\cal H}'_{\bm 0}  G_0)^{n-r-1} \right] 
        (\sigma^\beta G_0)  \,\biggr]  
\nonumber \\
&= \frac{1}{(n-2)!} \left( \frac{\partial}{\partial \zeta} \right)^{n-2}  
\nonumber \\
&\  \times   {\rm tr} \biggl[\,  
        \frac{1}{1-\zeta {\cal H}'_{\bm 0}  G_0}
         \sigma^\alpha  \frac{\partial}{\partial k_i}  
         \left[ G_0  \frac{1}{1-\zeta {\cal H}'_{\bm 0}  G_0} \right] 
        (\sigma^\beta G_0)  \,\biggr] \Biggr|_{\zeta=0}  
\nonumber \\
&=   \frac{1}{(n-2)!} \left( \frac{\partial}{\partial \zeta} \right)^{n-2}  
       {\rm tr} \biggl[\,  
          G_\zeta 
         \sigma^\alpha  \frac{\partial G_\zeta}{\partial k_i}  
         \sigma^\beta   \,\biggr] \Biggr|_{\zeta=0} . 
\end{align}
 In the last expression, we introduced   
\begin{align}
  G_\zeta &=  G_0 \frac{1}{1-\zeta {\cal H}'_{\bm 0} G_0 } 
= \frac{1}{i\varepsilon_m + \mu - {\cal H}_{\bm k}^0 - \zeta {\bm M} \cdot {\bm \sigma}}    , 
\end{align}
with uniform ${\bm M}$. 
 To take the $n$-sum, we note that it has the form, 
\begin{align}
  Q 
&\equiv \sum_{n=0}^\infty \frac{1}{n+2}  \frac{1}{n!} 
            \left( \frac{\partial}{\partial \zeta} \right)^n  P(\zeta) \Biggr|_{\zeta=0} . 
\end{align}
 Expanding $P(\zeta)$ in powers of $\zeta$,  
$\displaystyle  P(\zeta) = \sum_{n=0}^\infty p_n \zeta^n$,  
we find  
\begin{align}
  Q  &= \sum_{n=0}^\infty \frac{p_n}{n+2} = \int_0^1 \zeta d\zeta  P(\zeta)   . 
\end{align}
 Therefore, we obtain 
\begin{align}
 F &=  \sum_{\bm q}  (- q_i) M_{\bm q}^\alpha  M_{-{\bm q}}^\beta  
      \int_0^1 \zeta d\zeta \,  T \sum_l  \sum_{\bm k}  
       {\rm tr} \biggl[\,   G_\zeta 
         \sigma^\alpha  \frac{\partial G_\zeta}{\partial k_i}  
         \sigma^\beta   \,\biggr]   . 
\end{align}
 Integrating by parts with respect to $k_i$, 
we see that the coefficient of $M^\alpha  \partial_i M^\beta$ is antisymmetric 
with respect to $\alpha$ and $\beta$. 
 Therefore, $F$ can be written in the form, 
\begin{align}
 F &=   \int d^3 x \, D_i^\alpha ({\bm n} \times \partial_i {\bm n})^\alpha , 
\end{align}
and the coefficient is given by
\begin{align}
 D_i^\alpha &= \frac{1}{2i} \varepsilon_{\alpha\beta\gamma}  M^2 
        \int_0^1 \zeta \, d\zeta \, T\sum_m \sum_{\bm k}  
        {\rm tr} \left[ \sigma^\beta \frac{\partial G_\zeta}{\partial k_i} \sigma^\gamma G_\zeta  \right] . 
\label{eq:app_D0}
\end{align}
 If we regard $\zeta M \to M$ is an integration variable, we can also write 
\begin{align}
 D_i^\alpha &= \frac{1}{2i} \varepsilon_{\alpha\beta\gamma}   
        \int_0^M M dM \, T\sum_m \sum_{\bm k}  
        {\rm tr} \left[ \sigma^\beta \frac{\partial G_M}{\partial k_i} \sigma^\gamma G_M  \right] , 
\label{eq:app_D1}
\end{align}
where the integration is done with respect to the magnitude, $M = |{\bm M}|$, 
while keeping the direction ${\bm n} = {\bm M}/M$ constant.

\subsection{On spatial dependence of ${\bm M}$ in ${\bm D}_i ({\bm M})$ } 
\label{functional}

 As mentioned near the beginning of Sec.~\ref{Variations}, 
one can relax the condition of spatial uniformity of ${\bm n}={\bm M}/M$ 
on which ${\bm D}_i$ depends. 
 To show this, one may proceed in the same way as in the preceding subsection 
using as the unperturbed Green's function 
\begin{align}
 \hat G_M (i\varepsilon_m) 
&=  ( i\varepsilon_m + \mu - \hat {\cal H}^0 - M {\bm n} \cdot {\bm \sigma})^{-1}  , 
\label{eq:G_zeta}
\end{align}
with spatially nonuniform ${\bm n}$. 
 In the ${\bm k}$-representation, this $\hat G_M$ has off-diagonal components. 
 Treating $\hat G_M $ as a matrix in an infinite-dimensional Hilbert space, 
we can repeat the same manipulation and obtain 
\begin{align}
 D_i^\alpha &= \frac{1}{2i} \varepsilon_{\alpha\beta\gamma}   
        \int_0^M M dM \, T\sum_m  
        {\rm Tr} \left[ \sigma^\beta \hat G_{M} \hat v_i  \hat G_{M} \sigma^\gamma \hat G_{M}  \right] , 
\label{eq:result_2}
\end{align}
where $\hat v_i$ is the velocity operator, and Tr means the trace over all single-particle states 
(which generalizes the sum over ${\bm k}$, spin, and band). 
 This corresponds to retaining all the ${\bm q}_r$'s in the preceding subsection, 
without recourse to the approximation (\ref{eq:App:uniform_M}).

\section{Calculation of free energy variations}
\label{comparison}

  In this Appendix, we calculate first- and second-order variations of the DM free energy functional,  
\begin{align}
 F_{\rm DM} &= \int d{\bm r} \,  \tilde {\bm D}_i ({\bm M}) \cdot ({\bm M} \times \partial_i {\bm M}) ,  
\label{app:eq:F}
\end{align} 
under a small change ${\bm M} \to {\bm M} + \delta {\bm M}$. 
 Note that $\tilde {\bm D}_i = \tilde {\bm D}_i ({\bm M})$ is in general a function of ${\bm M} = {\bm M}({\bm r})$.  
 In the following calculation, it is convenient to define 
\begin{align}
 {\bm B}_i &=  \tilde {\bm D}_i ({\bm M}) \times {\bm M} , 
\end{align} 
and write 
\begin{align}
 F_{\rm DM} &= \int d{\bm r} \, {\bm B}_i \cdot \partial_i {\bm M} . 
\end{align}

\subsection{First-order variation} 
\label{d^1F}

 We start with the first-order variation, 
\begin{align}
 \delta^{(1)} F_{\rm DM} &= \int d{\bm r} \, 
   ( \delta {\bm B}_i \cdot \partial_i {\bm M} +  {\bm B}_i \cdot \partial_i \delta {\bm M} ) , 
\end{align} 
where $\delta {\bm B}_i = (\partial {\bm B}_i / \partial M^\alpha ) \delta M^\alpha$. 
 Integrating the second term by parts, dropping the surface term, and using 
$\partial_i {\bm B}_i = (\partial {\bm B}_i / \partial M^\beta) (\partial_i M^\beta)$, we have
\begin{align}
  \delta^{(1)} F_{\rm DM} 
&= \int d{\bm r}  
    \left( \frac{\partial B_i^\beta}{\partial M^\alpha} - \frac{\partial B_i^\alpha}{\partial M^\beta} \right) 
    \delta M^\alpha \partial_i M^\beta  
\nonumber \\
&\equiv \int d{\bm r}  \, \tilde {\bm D}_i^{(1)}  \cdot (\delta {\bm M} \times \partial_i {\bm M} ) , 
\label{eq:D1_def}  
\end{align} 
where we defined 
\begin{align}
 \tilde {\bm D}_i^{(1)}  &=  {\bm \nabla}_M \times {\bm B}_i  
 = {\bm \nabla}_M \times ( \tilde {\bm D}_i \times {\bm M} ) , 
\label{app:eq:D^1}
\end{align} 
with ${\bm \nabla}_M = \partial / \partial {\bm M} = (\partial / \partial M_x , \partial / \partial M_y ,\partial / \partial M_z)$.

\subsection{Second-order variation} 
\label{d^2F}

 The second-order variation is given by $\delta^{(2)} F_{\rm DM} =  F_1 + F_2$, 
\begin{align}
 F_1 &= \int d{\bm r} \,  \delta {\bm B}_i \cdot \partial_i \delta {\bm M}  , 
\label{eq:F1}
\\
 F_2 &= \int d{\bm r} \,  \delta^{(2)} {\bm B}_i \cdot \partial_i {\bm M}  , 
\label{eq:F2}
\end{align} 
where  $\delta {\bm B}_i = (\partial {\bm B}_i / \partial M^\alpha ) \, \delta M^\alpha$ (as above) 
and $\delta^{(2)} {\bm B}_i = \frac{1}{2} (\partial^2 {\bm B}_i /\partial M^\alpha \partial M^\beta ) 
       \, \delta M^\alpha \delta M^\beta $. 
 Integration by parts gives 
\begin{align}
 F_1 &=  - \int d{\bm r} \,  \frac{\partial B_i^\beta}{\partial M^\alpha} 
           (\partial_i \delta M^\alpha) \cdot  \delta M^\beta  
\nonumber \\
&\quad - \int d{\bm r} \,  \frac{\partial^2 B_i^\beta}{\partial M^\alpha \partial M^\gamma} 
           (\partial_i M^\gamma) \, \delta M^\alpha  \delta M^\beta  . 
\label{eq:C1b}
\end{align} 
 Summing Eqs.~(\ref{eq:F1}) and (\ref{eq:C1b}) leads to 
\begin{align}
 F_1
&=  \frac{1}{2}  \int d{\bm r} \, \left( 
      \frac{\partial B_i^\beta}{\partial M^\alpha} -  \frac{\partial B_i^\alpha}{\partial M^\beta}  \right)   
      \delta M^\alpha \, \partial_i \delta M^\beta  
\nonumber \\
&\quad - \frac{1}{2} \int d{\bm r} \,  \frac{\partial^2 B_i^\beta}{\partial M^\alpha \partial M^\gamma} 
           (\partial_i M^\gamma) \, \delta M^\alpha  \delta M^\beta . 
\label{eq:C1c}
\end{align} 
 We write the first term as
\begin{align}
  \delta^{(2a)} F_{\rm DM} 
&= \int d{\bm r}  \, \tilde {\bm D}_i^{(2)} \cdot (\delta {\bm M} \times \partial_i \delta {\bm M} ) , 
\label{eq:D2_def}
\end{align}  
where
\begin{align}
 2 \tilde {\bm D}_i^{(2)}  &=  {\bm \nabla}_M \times {\bm B}_i  
 = {\bm \nabla}_M \times ( \tilde {\bm D}_i \times {\bm M} ) . 
\label{app:eq:D^2}
\end{align} 
 The remaining terms [$F_2$ and the second term in Eq.~(\ref{eq:C1c})] are combined into 
\begin{align}
 \delta^{(2b)} F_{\rm DM} 
&= \frac{1}{2}  \int d{\bm r} \frac{ \partial}{\partial M^\alpha} 
        \left(  \frac{ \partial B_i^\gamma}{\partial M^\beta} 
             -  \frac{\partial B_i^\beta}{\partial M^\gamma} \right) 
            \delta M^\alpha \delta M^\beta  \, \partial_i M^\gamma  
\nonumber \\
&=  \int d{\bm r} \, \frac{ \partial \tilde {\bm D}_i^{(2)}}{\partial M^\alpha} 
            \delta M^\alpha \!\cdot (\delta {\bm M} \times \partial_i {\bm M}) 
\nonumber \\
&\equiv  \int d{\bm r} \, \delta \tilde {\bm D}_i^{(2)} \!\cdot (\delta {\bm M} \times \partial_i {\bm M})  . 
\label{eq:D2b_def}
\end{align}

\subsection{General formula for $\tilde {\bm D}_i^{(2)}$}
\label{general_relations}

 As shown in the preceding subsection, for given ${\bm D}_i$ (or $\tilde {\bm D}_i$), 
one can calculate $\tilde {\bm D}_i^{(1)}$ and $\tilde {\bm D}_i^{(2)}$ as 
\begin{align}
 \tilde {\bm D}_i^{(1)} = 2 \tilde {\bm D}_i^{(2)} 
&= {\bm \nabla}_{\bm M} \times (\tilde {\bm D}_i  \times {\bm M}) ,  
\label{eq:app_D^2_eq}
\end{align}
or
\begin{align}
 \tilde {\bm D}_i^{(1)} = 2 \tilde {\bm D}_i^{(2)} 
&= 2 \tilde {\bm D}_i + M \frac{\partial \tilde {\bm D}_i }{\partial M} - {\bm M} {\rm div}_M \tilde {\bm D}_i 
\nonumber \\
&= \frac{1}{M} \frac{\partial {\bm D}_i }{\partial M} - {\bm M} {\rm div}_M \tilde {\bm D}_i  , 
\label{eq:D_eq_2}
\end{align}
where ${\rm div}_M \tilde {\bm D}_i \equiv {\bm \nabla}_M \!\cdot\! \tilde {\bm D}_i$ 
and ${\bm D}_i = M^2 \tilde {\bm D}_i$. 
 These relations hold generally among $\tilde {\bm D}_i$, $\tilde {\bm D}_i^{(1)}$ 
and $\tilde {\bm D}_i^{(2)}$, which are  
defined by Eqs.~(\ref{app:eq:F}), (\ref{eq:D1_def}), and (\ref{eq:D2_def}) [and (\ref{eq:D2b_def})], respectively. 
 If the explicit expression for ${\bm D}_i$, Eq.~(\ref{eq:app_D1}), is used, 
one can prove (see the next subsection) 
\begin{align}
  {\rm div}_M  \tilde {\bm D}_i = 0 , 
\label{app:eq:divD=0}
\end{align}
and Eq.~(\ref{eq:D_eq_2}) reduces to  
\begin{align}
   \tilde {\bm D}_i^{(1)}  = 2 \tilde {\bm D}_i^{(2)} &= \frac{1}{M} \frac{\partial {\bm D}_i}{\partial M}  . 
\end{align}
 Thus, we find a general formula for $\tilde {\bm D}_i^{(2)}$, 
\begin{align}
  \tilde D_i^{(2), \alpha} &= \frac{1}{4i} \varepsilon_{\alpha\beta\gamma}   
         \, T\sum_m \sum_{\bm k}  
        {\rm tr} \left[ \sigma^\beta \frac{\partial G_M}{\partial k_i} \sigma^\gamma G_M  \right] , 
\end{align} 
which does not involve the $M$- (or $\zeta$-) integration. 
 Conversely, if $\tilde {\bm D}_i^{(2)}$ is given, ${\bm D}_i$ can be obtained by integration, 
\begin{align}
 {\bm D}_i  ({\bm M}) &= 2 \int_0^M M' dM'  \tilde {\bm D}_i^{(2)} ({\bm M}') , 
\label{eq:app_D_from_D2}
\end{align} 
where ${\bm M} = M {\bm n}$ and ${\bm M}' = M' {\bm n}$. 
 The integration is done with respect to the magnitude of ${\bm M} $, with the direction ${\bm n}$ kept fixed.

\subsection{Proof of ${\rm div}_M \tilde {\bm D}_i = 0$}
\label{proof}

 To prove ${\rm div}_M \tilde {\bm D}_i = 0$, we use Eq.~(\ref{eq:app_D0}) and write  
\begin{align}
 \tilde D_i^\alpha &=  \frac{1}{2i} \varepsilon_{\alpha\beta\gamma}  
       \int_0^1 \zeta \, d\zeta \, T\sum_m \sum_{\bm k}  
      {\rm tr} \left[ \sigma^\beta \frac{\partial G_\zeta}{\partial k_i} \sigma^\gamma G_\zeta  \right] . 
\label{eq:app_D3}
\end{align}
 Since only the integrand, 
\begin{align}
 d_i^\alpha ({\bm M}) 
&\equiv  \varepsilon_{\alpha\beta\gamma}  
     \sum_{\bm k}  
   {\rm tr} \left[ \sigma^\beta  G_\zeta v_i G_\zeta \sigma^\gamma G_\zeta  \right] , 
\label{eq:app_d_M}
\end{align}
depends on ${\bm M}$, it is sufficient to prove 
\begin{align}
 {\rm div}_M \, {\bm d}_i ({\bm M}) = 0 . 
\label{eq:app_div=0}
\end{align}
 Using $\partial G_\zeta / \partial M^\alpha = \zeta \, G_\zeta \sigma^\alpha G_\zeta$, one has 
\begin{align}
 {\rm div}_M \, {\bm d}_i 
&=  \varepsilon_{\alpha\beta\gamma} \, \zeta 
    \sum_{\bm k}   
    {\rm tr} \left[ \sigma^\beta G_\zeta \sigma^\alpha G_\zeta v_i G_\zeta \sigma^\gamma G_\zeta  \right] 
\nonumber \\
& \hskip 18mm 
        + (\text{2 similar terms})    , 
\label{eq:app_div=0_2}
\end{align}
which consists of terms of the form, 
\begin{align}
 I &\equiv  \varepsilon_{\alpha\beta\gamma}   
        \sum_{\bm k}   
        {\rm tr} \left[  G_\zeta v_i G_\zeta \sigma^\gamma G_\zeta  \sigma^\beta G_\zeta \sigma^\alpha \right] . 
\end{align}
 Noting $G_\zeta v_i G_\zeta = \partial G_\zeta / \partial k_i$ and integrating by parts, we find 
\begin{align}
 I &=  - \varepsilon_{\alpha\beta\gamma}   
        \sum_{\bm k} \biggl\{ 
        {\rm tr} \left[  G_\zeta \sigma^\gamma G_\zeta v_i G_\zeta  \sigma^\beta G_\zeta \sigma^\alpha \right] 
\nonumber \\
& \hskip 18mm 
        +  {\rm tr} \left[  G_\zeta \sigma^\gamma G_\zeta  \sigma^\beta G_\zeta v_i G_\zeta \sigma^\alpha \right] 
  \biggr\}  
\nonumber \\
&=  -2I , 
\end{align}
hence, $I = 0$. 
 This means that the right-hand side of Eq.~(\ref{eq:app_div=0_2}) vanishes, 
proving Eq.~(\ref{eq:app_div=0}), thus ${\rm div}_M \tilde {\bm D}_i = 0$.

\subsection{Other consequences of ${\rm div}_M \tilde {\bm D}_i = 0$}
\label{Remarks}

 The condition ${\rm div}_M \tilde {\bm D}_i = 0$ also provides some insight into the connection between 
the functional form (i.e., dependence on ${\bm M}$) and the tensorial structure 
(i.e., dependence on $i$ and $\alpha$) of $\tilde D_i^\alpha$.

 If $\tilde {\bm D}_i$ has the form,  $\tilde D_i^\alpha = \tilde C_1 \, \delta_{i\alpha}$, 
in 3D, where $i, \alpha = x,y,z$, 
the above condition says that $\tilde D_i^\alpha$ does not depend on ${\bm M}$. 
 It then follows from Eq.~(\ref{eq:D^2_eq1}) that $\tilde {\bm D}_i^{(2)} = \tilde {\bm D}_i$. 
 This occurs in magnetic Weyl semimetals, as we have seen in Sec.~III B.

 If $\tilde {\bm D}_i$ has the form, $\tilde D_i^\alpha = \tilde C_1 \, \varepsilon_{i\alpha}$, 
where $\varepsilon_{i\alpha}$ is the antisymmetric tensor in 2D, 
$\tilde {\bm D}_i$ does not depend on the in-plane components, $M_x$ and $M_y$. 
 This is the case for topological insulator surface states.  
 Indeed, $\tilde D_i^\alpha$ depends only on $M_z$, 
as shown in Ref.~\onlinecite{Wakatsuki2015} and also in Sec.~III C.

 Unfortunately, such analysis is not effective for a Rashba ferromagnet, 
in which $\tilde D_i^\alpha$ takes the form, 
$\tilde D_i^\alpha = \tilde C_1 \varepsilon_{i\alpha} + \tilde C_2 ({\bm n} \times \hat z)_i n^\alpha$. 
 While the $\tilde C_2$-term does not contribute to the DM free energy, its presence hinders the extraction 
of any information on $\tilde C_1$ from ${\rm div}_M \tilde {\bm D}_i = 0$. 
 Similar feature exists also in the weak SOC case, as studied in Appendix \ref{DMI_SOI1}. 

 These are summarized in Table I.

\section{Weak spin-orbit coupling}
\label{weak_SOC}

 In this section, we show that at the first order in SOC, the DMI coefficient (in free energy) is given 
by the equilibrium spin-current density. 
 Such relation was found in Ref.~\onlinecite{Kikuchi2016}, and here we make a slight generalization. 
 Since the SOC is treated perturbatively, topological materials are not eligible.

 We consider the Hamiltonian 
\begin{align}
 H = \sum_{\bm k} c^\dagger_{\bm k} (\varepsilon_{\bm k}
   + \lambda_i^\alpha \gamma_i ({\bm k}) \sigma^\alpha  
    + \bm{M} \cdot \bm{\sigma}  ) c_{\bm k} , 
\label{eq:H_weak_soc}
\end{align}
where $c_{\bm k} = {}^t (c_{\bm{k}\uparrow}, c_{\bm{k} \downarrow})$. 
 The second term describes the SOC with strength $\lambda_i^\alpha$. 
 A free electron model would assume $\varepsilon_{\bm k} = k^2/2m$ and $\gamma_i ({\bm k}) = k_i$, 
as in Ref.~\onlinecite{Kikuchi2016}, 
but here, we leave $\varepsilon_{\bm k}$ and $\gamma_i ({\bm k})$ to be arbitrary functions 
of ${\bm k}$. 
 As shown below, even with this setting, the DMI coefficient is still given 
by the equilibrium spin current at the first order in $\lambda_i^\alpha$.

 The full Green's function is given by
\begin{align}
 G_{M,\lambda} = (g_0-\bm{M} \cdot \bm{\sigma} - \lambda_i^\alpha \gamma_i ({\bm k}) \sigma^\alpha)^{-1}.
\end{align}
with $g_0 = i\varepsilon_m + \mu - \varepsilon_{\bm k}$.
 We expand it up to the first order in $\lambda_i^\alpha$,  
$ G_{M,\lambda} \simeq G_{M,0} + \lambda_i^\eta \gamma_i ({\bm k}) G_{M,0} \sigma^\eta G_{M,0}$, 
where 
\begin{align}
\label{Green_function_M}
 G_{M,0}  = \frac{g_0+\bm{M}\cdot \bm{\sigma} }{D_{M,0}},
\end{align}
is $G_{M,\lambda}$ at $\lambda_i^\alpha = 0$, with $D_{M,0}= g_0^2 -M^2$.

\subsection{DMI coefficient}
\label{DMI_SOI1}

 The DMI coefficient (\ref{eq:result}) at the first-order in $\lambda_i^\alpha$ is then 
\begin{align}
 D_i^\alpha 
&= - i \lambda_j^\eta \varepsilon_{\alpha \beta \gamma} \int^M_0 M dM \, 
     T \sum_{n, {\bm k}}  v_i^0 \gamma_j ({\bm k})
\nonumber \\
&\hspace{40pt} \times {\rm tr } \left[ \sigma^\beta  (G_{M,0})^2 \sigma^\gamma G_{M,0} \sigma^\eta G_{M,0} \right] , 
\end{align}
where we used $\partial G_{M,0} /\partial k_i = v_i^0 (G_{M,0})^2$ 
with $v_i^0 = \partial \varepsilon_{\bm k}/\partial k_i$. 
 The spin trace is taken as 
\begin{align}
& \varepsilon_{\alpha \beta \gamma} {\rm tr } 
 \left[ \sigma^\beta  (G_{M,0})^2 \sigma^\gamma G_{M,0} \sigma^\eta G_{M,0} \right] 
\nonumber \\
& =  4i  \frac{(g_0^2+M^2) \delta_{\alpha \eta} - 2M^\alpha M^\eta}{D_{M,0}^3}  
\nonumber \\
&   =  \frac{2i}{M} \frac{d}{dM} \biggl\{ \frac{M^2 (\delta_{\alpha \eta} - n^\alpha n^\eta) }{D_{M,0}^2}    
           + \frac{n^\alpha n^\eta}{D_{M,0}}  \biggr\} .  
\label{eq:weak_soc_calc}
\end{align}
 Since the ${\bm M}$-parallel component of ${\bm D}_i$ does not contribute to the DM free energy, 
one may drop the second term ($\sim n^\alpha n^\eta$) and thereby define ${\bm D}_i^\perp$.  
 The $M$-integration can be done analytically, 
\begin{align}
 \bm{D}_i^\perp  
&= 2\bm{\lambda}_j^\perp T \sum_{n, {\bm k}} v_i^0 \gamma_j ({\bm k}) 
     \int^M_0 dM  \frac{d}{dM} \frac{M^2}{D_{M,0}^2} 
\nonumber\\
&= 2\bm{\lambda}_j^\perp T \sum_{n, {\bm k}} v_i^0 \gamma_j ({\bm k}) 
       \frac{M^2}{D_{M,0}^2} 
\\
&= - \bm{\lambda}_j^\perp T \sum_{n, {\bm k}}  v_i^0 \gamma_j ({\bm k}) 
      \left[\frac{g_0^2-M^2}{D_{M,0}^2} - \frac{g_0^2+M^2}{D_{M,0}^2} \right] 
\nonumber\\
&= - \bm{\lambda}_j^\perp T \sum_{n, {\bm k}}  
      \left[ \frac{v_i^0 \gamma_j ({\bm k})}{D_{M,0}} + (\partial_i \gamma_j ) \frac{g_0}{D_{M,0}}  \right] , 
\label{eq:D_weak}
\end{align}
where 
${\bm \lambda}_i^\perp \equiv {\bm \lambda}_i - ({\bm \lambda}_i \cdot {\bm n}) \, {\bm n}$ 
is the component perpendicular to ${\bm M}$. 
 In the last equality, we noted 
\begin{align}
 \frac{g_0^2 + M^2}{D_{M,0}^2} v_i^0 &= \frac{\partial}{\partial k_i} \frac{g_0}{D_{M,0}}  , 
\label{eq:note1}
\end{align}
and made an integration by parts.

 As to the parallel component of ${\bm D}_i^{(2)}$,  Eq.~(\ref{eq:weak_soc_calc}) leads to 
\begin{align}
  {\bm D}_i^{(2), \parallel} 
&= {\bm \lambda}_j^\parallel  M \frac{d}{dM} 
     T \sum_{n, {\bm k}} \frac{v_i^0 \gamma_j ({\bm k})}{D_{M,0}} 
\nonumber \\
&= 2 {\bm \lambda}_j^\parallel  M^2 
     T \sum_{n, {\bm k}} \frac{v_i^0 \gamma_j ({\bm k})}{D_{M,0}^2} . 
\label{eq:D2_weak}
\end{align}
 This is consistent with the analysis summarized in Table I.

\subsection{Equilibrium spin-current density}
\label{spin-current}

 Next, we express the equilibrium spin current in terms of Green's functions.  
 The spin current is expressed as
\begin{align}
 j_{s,i}^\alpha 
= \frac{1}{4} \sum_{\bm k} \langle c_{\bm k}^\dagger (v_i\sigma^\alpha +\sigma^\alpha v_i ) c_{\bm k} \rangle  , 
\label{eq:app:js}
\end{align}
where $v_i = v_i^0 + \lambda_j^\beta (\partial_i \gamma_j ) \, \sigma^\beta$.
 Thus,
\begin{align}
\label{jsi1}
  j_{s,i}^\alpha 
&= \frac{1}{2} T\sum_{n,\bm{k}}  v_i^0 \mathrm{tr} \left[ \sigma^\alpha G_{M,\lambda} \right] 
   + \frac{1}{2} T \sum_{n,\bm{k}}  \lambda^\alpha_j (\partial_i \gamma_j ) \mathrm{tr} \left[ G_{M,\lambda} \right]  
\nonumber \\
&\simeq  \frac{1}{2} \lambda_j^\beta  T\sum_{n,\bm{k}} v_i^0 \gamma_j ({\bm k})
        \mathrm{tr} \left[ \sigma^\alpha G_{M,0} \sigma^\beta G_{M,0} \right] 
\nonumber \\
&\quad   + \frac{1}{2} \lambda_j^\alpha T \sum_{n,\bm{k}} (\partial_i \gamma_j ) \mathrm{tr} \left[ G_{M,0} \right] .
\end{align}
 In the second line, we retained only the first-order terms in $\lambda_i^\alpha$. 
 The spin trace is calculated as  
\begin{align}
\mathrm{tr} \left[ \sigma^\alpha G_{M,0}  \sigma^\beta G_{M,0} \right] 
=  \frac{2(g_0^2 -M^2) \delta_{\alpha\beta} + 4 M^\alpha M^\beta}{D_{M,0}^2} .  
\end{align}
 We first note that the parallel component vanishes, 
\begin{align}
 {\bm n} \cdot {\bm j}_{s,i} &= \bm{\lambda}_j^\parallel  T\sum_{n,\bm{k}} 
     \left[ v_i^0 \gamma_j ({\bm k}) \frac{g_0^2 + M^2}{D_{M,0}^2} + (\partial_i \gamma_j ) \frac{g_0}{D_{M,0}}  \right] 
\nonumber \\
&= 0 ,  
\end{align}
because of Eq.~(\ref{eq:note1}). 
 Thus, the spin current consists purely of the perpendicular component, 
\begin{align}
 \bm{j}_{s,i}
&= \bm{\lambda}_j^\perp  T\sum_{n,\bm{k}} 
    \left[  \frac{v_i^0 \gamma_j ({\bm k}) }{D_{M,0}} + (\partial_i \gamma_j )  \frac{g_0}{D_{M,0}} \right] .  
\end{align}
 Comparing this with Eq.~(\ref{eq:D_weak}), we see that  
\begin{align}
 \bm{D}_i^\perp  = - {\bm j}_{s,i}  ,  
\end{align}
confirming that our general formula reduces to the spin-current formula of Ref.~\onlinecite{Kikuchi2016} 
at weak SOC. 
 We emphasize that this holds even when $\varepsilon_{\bm k}$ and $\gamma_i ({\bm k})$ 
are arbitrary functions of ${\bm k}$, 
which was not obvious from the derivation in Ref.~\onlinecite{Kikuchi2016} .

\subsection{Explicit evaluation}
\label{Specific_value}

 Finally, we evaluate the results explicitly for free electrons in $d$ dimensions. 
 We start with general $\varepsilon_{\bm k}$ and $\gamma_i ({\bm k})$, and then specialize in stages 
to $\varepsilon_{\bm k} = {\bm k}^2/2m$ and $\gamma_j ({\bm k}) = k_j$.

 Defining $G_{\uparrow} \equiv (g_0 + M)^{-1}$ and $G_{\downarrow} \equiv (g_0 - M)^{-1}$, and using 
$ 2MD_{M,0}^{-1} =G_{\downarrow} -  G_{\uparrow}$ and $ 2g_0D_M^{-1} = G_{\uparrow} + G_{\downarrow}$, 
Eq.~(\ref{eq:D_weak}) is calculated as follows, 
\begin{align}
 {\bm D}_i^\perp  
&= - \bm{\lambda}_j^\perp T \sum_{n, {\bm k}}  
 \left[ - v_i^0 \gamma_j ({\bm k}) \frac{G_{\uparrow} - G_{\downarrow}}{2M}   
 + (\partial_i \gamma_j ) \frac{G_{\uparrow} + G_{\downarrow}}{2} \right]  
\nonumber \\
&= - \bm{\lambda}_j^\perp \sum_{\bm k}  
 \left[ - v_i^0 \gamma_j ({\bm k}) \frac{f_{\uparrow} - f_{\downarrow}}{2M}   
 + (\partial_i \gamma_j ) \frac{f_{\uparrow} + f_{\downarrow}}{2} \right]  
\nonumber \\
&= \bm{\lambda}_j^\perp \sum_{\bm k}  v_i^0 \gamma_j ({\bm k}) 
 \left[ \frac{f_{\uparrow} - f_{\downarrow}}{2M}   
 + \frac{f'_{\uparrow} + f'_{\downarrow}}{2} \right]  , 
\label{eq:D_weak_soc_1}
\end{align}
where $f_\sigma = f (\varepsilon_{\bm k} - \sigma M)$ is the Fermi distribution function 
and $f'_\sigma = \partial f_\sigma / \partial \varepsilon$. 
 The last expression shows that ${\bm D}_i^\perp$ is proportional to $M^2$ 
at small $M$ ($\ll \varepsilon_{\rm F}$). 
 This fact was pointed out in Ref.~\onlinecite{Freimuth2017} in a different formulation. 
 These hold for general $\varepsilon_{\bm k}$ and $\gamma_j ({\bm k})$.

 For $\gamma_j ({\bm k}) = k_j$ (but with general $\varepsilon_{\bm k}$), 
${\bm D}_i^\perp$ is expressed as 
\begin{align}
  {\bm D}_i^\perp 
&= - {\bm \lambda}_i^\perp 
  \biggl[ \frac{\Omega_{\uparrow} - \Omega_{\downarrow}}{2M} + \frac{n}{2}  \biggr] , 
\label{eq:D_weak_soc_2}
\end{align}
where 
$\Omega_\sigma =  - T\sum_{\bm k} \ln (1 + e^{-\beta (\varepsilon_{\bm k} - \sigma M - \mu)})$ 
is the thermodynamic potential of spin-$\sigma$ electrons, and   
$n = n_{\uparrow}+n_{\downarrow}$ is the total electron density. 
 The full ${\bm D}_i$ and ${\bm D}_i^{(2)} $ are obtained as 
\begin{align}
  {\bm D}_i = C_1 {\bm \lambda}_i^\perp + C_2 {\bm \lambda}_i^\parallel ,  
\label{eq:D_weak_soc}
\\
 {\bm D}_i^{(2)} = C_1 {\bm \lambda}_i^\parallel  +  C_3 {\bm \lambda}_i^\perp ,
\label{eq:D2_weak_soc}
\end{align}
with 
\begin{align}
  C_1 &= - \frac{\Omega_\uparrow - \Omega_\downarrow}{2M} - \frac{n}{2} ,  
\label{eq:C1_weak_soc}
\\
  C_2 &= \frac{\Omega_\uparrow - \Omega_\downarrow}{M} + n(0)  ,
\label{eq:C2_weak_soc}
\\
  C_3 &= \frac{1}{2} \left[ C_1 + \frac{M}{2} (\bar\nu_\uparrow - \bar\nu_\downarrow) \right] , 
\label{eq:C3_weak_soc}
\end{align}
where $n(0)$ is the electron density at $M=0$ (for given $\mu$),  
and $\bar\nu_\sigma = \sum_{\bm k} (-f_\sigma') $ is the thermally-averaged density of states. 
 Here we used $C_1 = \frac{M}{2} \frac{\partial C_2}{\partial M}$ and 
$C_3 = \frac{M}{2} \frac{\partial C_1}{\partial M}$ as listed in Table \ref{table_I}.

 Finally, for $\gamma_j ({\bm k}) = k_j$ and $\varepsilon_{\bm k} = {\bm k}^2/2m$, 
$C_1$ becomes \cite{Ikeda2013}  
\begin{align}
  C_1  
&= \frac{ \varepsilon_{\mathrm{F} \uparrow} n_{\uparrow}-\varepsilon_{\mathrm{F}\downarrow} n_{\downarrow}  }{M(d+2)}  
  - \frac{n}{2}  , 
\label{eq:D_weak_soc_2}
\end{align}
at $T=0$, where $\varepsilon_{\mathrm{F} \sigma} = \varepsilon_{\rm F} + \sigma M$ with Fermi energy 
$\varepsilon_{\mathrm{F}}$. 
  For $M \ll \varepsilon_{\rm F}$, in which there are two Fermi surfaces, we have 
\begin{align}
 C_1
&=   - \frac{d-2}{6 \, \Gamma (\frac{d}{2})} 
     \left( \frac{m \varepsilon_{\rm F}}{2\pi \hbar^2} \right)^{d/2} 
      \left( \frac{M}{\varepsilon_{\rm F}} \right)^2  + O(M^4) , 
\label{eq:D_weak_soc_3}
\end{align}
where $\Gamma$ is the Gamma function. 
  For $M > \varepsilon_{\rm F}$, where there is only one Fermi surface, we have 
\begin{align}
 C_1
&=    \frac{1}{2 \,\Gamma (\frac{d}{2} + 2)} 
     \left( \frac{m (\varepsilon_{\rm F} + M)}{2\pi \hbar^2} \right)^{d/2} 
    \left( \frac{\varepsilon_{\rm F}}{M} - \frac{d}{2}  \right) . 
\label{eq:D_weak_soc_4}
\end{align}

\section{Generalized spin-current formula}
\label{spin_current_formula}

 In this Appendix, we show that the spin current formula for the DMI coefficient holds 
more generally, irrespective of the strength of SOC, 
if the magnetization vector is perpendicular to the spin-orbit field, 
${\bm M} \perp {\bm \lambda}_i$. 
 The derivation is based on the same model as in Appendix C, namely, Eq.~(\ref{eq:H_weak_soc}), 
but the SOC is fully taken into account.  
 The spin current operator is given by Eq.~(\ref{eq:app:js}), 
and the equilibrium spin current is expressed as 
\begin{align}
 {\bm j}_{{\rm s},i}  
&= - T \sum_m \sum_{\bm k} \frac{(\partial_i g_0) {\bm g}_M - g_0 (\partial_i {\bm g}_M)}{D_M} . 
\label{eq:spin_current}
\end{align}

\subsection{Derivation}

 We first recall that the DMI coefficient is given by Eq.~(\ref{eq:D_2x2}), or its $M$-integration, 
\begin{equation}
  \tilde {\bm D}_i 
= \int_0^M 2MdM T \sum_m \sum_{\bm k} 
   \frac{(\partial_i g_0) {\bm g}_M - g_0 (\partial_i {\bm g}_M)}{(g_0^2 - {\bm g}_M^2)^2} , 
\end{equation}
 Assuming ${\bm M}$ is orthogonal to ${\bm \lambda}_i$, we focus on the perpendicular component, 
\begin{align}
  {\bm D}_i^\perp 
= \int_0^M 2MdM T \sum_m \sum_{\bm k} 
  \frac{(\partial_i g_0) {\bm g}_\perp - g_0 (\partial_i {\bm g}_\perp)}{(g_0^2 - {\bm g}_\perp^2 - M^2)^2} , 
\label{eq:D_2x2_1}
\end{align}
where ${\bm g}_\perp = {\bm g}_M - {\bm M} = {\bm \lambda}_i \gamma_i ({\bm k})$, 
and we noted ${\bm g}_M^2 = {\bm g}_\perp^2 + M^2$. 
 Since $M$ is not contained in $g_0$ and ${\bm g}_\perp$, 
the $M$-integration can be done exactly, 
\begin{align}
  {\bm D}_i^\perp  
&= \int_0^{M^2} dM^2   \, T \sum_m \sum_{\bm k} 
  \frac{(\partial_i g_0) {\bm g}_\perp - g_0 (\partial_i {\bm g}_\perp)}{(g_0^2 - {\bm g}_\perp^2 - M^2)^2} 
\nonumber \\
&= T \sum_m \sum_{\bm k} 
  \frac{(\partial_i g_0) {\bm g}_\perp - g_0 (\partial_i {\bm g}_\perp)}{g_0^2 - {\bm g}_\perp^2 - M^2} 
\nonumber \\
&\quad - T \sum_m \sum_{\bm k} 
  \frac{(\partial_i g_0) {\bm g}_\perp - g_0 (\partial_i {\bm g}_\perp)}{g_0^2 - {\bm g}_\perp^2}  .
\end{align}
 The resulting quantity is identified as the perpendicular component of the equilibrium spin current 
${\bm j}_{{\rm s},i}$ in Eq.~(\ref{eq:spin_current}). 
 Therefore, we obtain 
\begin{align}
  {\bm D}_i^\perp  &= - \{ {\bm j}_{{\rm s},i}^\perp (M) - {\bm j}_{{\rm s},i}^\perp (0) \} . 
\label{eq:app:D_spin_current}
\end{align}

\subsection{Application to Rashba ferromagnet}
\label{Rashba_spin_current}

 Let us consider the Rashba ferromagnet studied in Sec.~\ref{Rashba}. 
 The equilibrium spin current is expressed as 
\begin{align}
 j_{{\rm s},i}^{\perp, \alpha} 
&= \frac{\alpha_{\rm R}}{2} \varepsilon_{\alpha i} \sum_{\eta = \pm 1}     
  \sum_{\bm k}  \left(  1 - \eta \frac{v_x^0 k_x}{g_k} \right)  
     f ( \xi_k - \eta g_k )  . 
\end{align}
 This can be calculated analytically as follows, 
\begin{align}
  1) \  & \mu > M  :  
\nonumber \\ 
& \qquad 
  j_{{\rm s},i}^{\perp, \alpha}  (M) = \varepsilon_{\alpha i} \frac{m^2 \alpha_{\rm R}^3}{6 \pi}   , 
\\
  2) \  & -M < \mu < M : 
\nonumber \\
  & j_{{\rm s},i}^{\perp, \alpha}  (M) = \varepsilon_{\alpha i} \frac{m^2 \alpha_{\rm R}^3}{12 \pi}  
  \left\{ 1 - (\beta M)^3  + (P^2 - 3\beta \mu ) P \right\} , 
\\
  3) \  & {\cal E}_M < \mu < - M \ \ (\text{in case of }\beta M < 1) : 
\nonumber \\
& \qquad  
    j_{{\rm s},i}^{\perp, \alpha}  (M) = \varepsilon_{\alpha i} \frac{m^2 \alpha_{\rm R}^3}{6 \pi}  
   (P^2 - 3 \beta \mu ) P  , 
\end{align}
where 
\begin{align}
 \beta &= \hbar^2/m \alpha_{\rm R}^2 , 
\\
  P &= \sqrt{1 + 2 \beta \mu + (\beta M)^2} , 
\\
  {\cal E}_M &= - \frac{1}{2} \left( \frac{1}{\beta} + \beta M^2 \right) . 
\end{align}
 At $M=0$, it becomes 
\begin{align}
  {\rm a)} \  & \mu > 0  : 
\nonumber \\
&\qquad 
   j_{{\rm s},i}^{\perp, \alpha}  (0) = \varepsilon_{\alpha i} \frac{m^2 \alpha_{\rm R}^3}{6 \pi}   , 
\\
  {\rm b)} \  & {\cal E}_0 < \mu <  0 : 
\nonumber \\
  &\qquad  j_{{\rm s},i}^{\perp, \alpha}  (0) = \varepsilon_{\alpha i} \frac{m^2 \alpha_{\rm R}^3}{6 \pi}  
   (1 - \beta \mu ) P_0  , 
\\
  {\rm c)} \  &  \mu < {\cal E}_0 :  \quad  
  j_{{\rm s},i}^{\perp, \alpha}  (0) = 0 , 
\end{align}
where 
\begin{align}
 P_0 &= \sqrt{1 + 2 \beta \mu} , 
\\
 {\cal E}_0 &= - \frac{1}{2\beta} . 
\end{align}
 These lead to the DMI coefficient, Eq.~(\ref{eq:app:D_spin_current}), as follows. 

\ 

\vskip 3mm
\noindent
$\bullet$ \underline{Strong SOC} ($\beta M < \frac{1}{2}$ or ${\cal E}_0 < -M$)

\vskip 2mm

\begin{align}
  {\rm 1a)} \  & \mu > M  : \quad 
    D_i^{\perp, \alpha}  = 0   , 
\\
  {\rm 2a)} \  & 0 < \mu < M : 
\nonumber \\
  & D_i^{\perp, \alpha}  = \varepsilon_{i \alpha} \frac{m^2 \alpha_{\rm R}^3}{12 \pi}  
  \left\{  (P^2 - 3\beta \mu ) P  - (\beta M)^3 - 1   \right\} , 
\\
  {\rm 2b)} \  & -M < \mu < 0 : 
\nonumber \\
  & D_i^{\perp, \alpha}  = \varepsilon_{i \alpha} \frac{m^2 \alpha_{\rm R}^3}{12 \pi}  
  \bigl\{ (P^2 - 3\beta \mu ) P - 2 (1 - \beta \mu ) P_0 
\nonumber \\
  &\qquad\qquad\qquad\quad 
    - (\beta M)^3  +  1  \bigr\}  
\\
  {\rm 3b)} \  & {\cal E}_0 < \mu < - M \ \ (\text{in case of }\beta M < 1) : 
\nonumber \\
  & D_i^{\perp, \alpha}  = \varepsilon_{i \alpha} \frac{m^2 \alpha_{\rm R}^3}{6 \pi}  
    \left\{ (P^2 - 3 \beta \mu ) P -  (1 - \beta \mu ) P_0  \right\}  , 
\\
  {\rm 3c)} \  & {\cal E}_M < \mu < {\cal E}_0 \ \ (\text{in case of }\beta M < 1) : 
\nonumber \\
  & D_i^{\perp, \alpha}  = \varepsilon_{i \alpha} \frac{m^2 \alpha_{\rm R}^3}{6 \pi}  
   (P^2 - 3 \beta \mu ) P . 
\end{align}

\ 

\noindent
$\bullet$ \underline{Weak SOC} ($\beta M > \frac{1}{2}$ or ${\cal E}_0 > -M$)

\vskip 2mm

 In this case, 2b), 3b) and 3c) above are replaced by 
\begin{align}
  {\rm 2b)} \  &  {\cal E}_0 < \mu < 0 : 
\nonumber \\
  & D_i^{\perp, \alpha}    = \varepsilon_{i\alpha} \frac{m^2 \alpha_{\rm R}^3}{12 \pi}  
  \bigl\{ (P^2 - 3\beta \mu ) P  - 2 (1 - \beta \mu ) P_0 
\nonumber \\
  &\qquad\qquad\qquad\quad
    - (\beta M)^3   + 1   \bigr\}  , 
\\
  {\rm 2c)} \  & -M < \mu <  {\cal E}_0 : 
\nonumber \\
  & D_i^{\perp, \alpha}   = \varepsilon_{i\alpha}  \frac{m^2 \alpha_{\rm R}^3}{12 \pi}  
  \left\{ (P^2 - 3\beta \mu ) P - (\beta M)^3 + 1  \right\} , 
\\
  {\rm 3c)} \  & {\cal E}_M < \mu < - M \ \ (\text{in case of }\beta M < 1) : 
\nonumber \\
  & D_i^{\perp, \alpha}    = \varepsilon_{i\alpha}  \frac{m^2 \alpha_{\rm R}^3}{6 \pi}  
   (P^2 - 3 \beta \mu ) P . 
\end{align}


\vfill


\begin{thebibliography}{}

\bibitem{Dzyaloshinsky} I. E. Dzyaloshinskii, Sov. Phys. JETP {\bf 5}, 1259 (1957). 

\bibitem{Moriya} T. Moriya, Phys. Rev. {\bf 120}, 91 (1960).

\bibitem{Dzyaloshinsky1964} I. E. Dzyaloshinskii, Sov. Phys. JETP {\bf 19}, 960 (1964). 

\bibitem{Thiaville2012} A. Thiaville, S. Rohart, \'E. Ju\'e, V. Cros, and A. Fert, Europhys. Lett. {\bf 100}, 57002 (2012).

\bibitem{Chen2013} G. Chen, T. Ma, A. T. N'Diaye, H. Kwon, C. Won, Y. Wu, and A. K. Schmid, Nat. Commun. {\bf 4}, 2671 (2013).

\bibitem{Ryu2013} K.-S. Ryu, L. Thomas, S.-H. Yang, and S. Parkin, Nat. Nanotechnol. {\bf 8}, 527 (2013).

\bibitem{Emori2013} S. Emori, U. Bauer, S.-M. Ahn, E. Martinez, and G. S. D. Beach, Nat. Mater. {\bf 12}, 611 (2013).

\bibitem{Torrejon2014} J. Torrejon, J. Kim, J. Sinha, S. Mitani, M. Hayashi, M. Yamanouchi, and H. Ohno, Nat. Commun. {\bf 5}, 4655 (2014).

\bibitem{Bogdanov1989} A. N. Bogdanov, and D. A. Yablonskii, Sov. Phys. JETP {\bf 68}, 101 (1989).

\bibitem{Rossler2006} U. K. R\"ossler, A. N. Bogdanov, and C. Peiderer, Nature(London) {\bf 442}, 797 (2006).

\bibitem{Muhlbauer2009} 
S. M\"uhlbauer, B. Binz, F. Jonietz, C. Pfleiderer, A. Rosch, A. Neubauer, R. Georgii, and P. B\"oni, Science {\bf 323}, 915 (2009).

\bibitem{Yu2010} X. Z. Yu, Y. Onose, N. Kanazawa, J. H. Park, J. H. Han, Y. Matsui, N. Nagaosa, and
Y. Tokura, Nature {\bf 465}, 901 (2010).

\bibitem{Heinze2011} S. Heinze, K. von Bergmann, M. Menzel, J. Brede, A. Kubetzka, R. Wiesendanger, G. Bihlmayer, and S. Bl\"ugel, Nat. Phys. {\bf 7}, 713 (2011).

\bibitem{Nagaosa2013} N. Nagaosa and Y. Tokura, Nat. Nanotechnol. {\bf 8}, 899 (2013).
\bibitem{Shirane1983}  G. Shirane, R. Cowley, C. Majkrzak, J. B. Sokoloff, B. Pagonis, C. H. Perry, and Y. Ishikawa, 
Phys. Rev. B {\bf 28}, 6251 (1983).

\bibitem{Ishimoto1986}  K. Ishimoto, Y. Yamaguchi, S. Mitsuda, M. Ishida, and Y. Endoh, 
J. Magn. Magn. Mater. {\bf 54} 1003 (1986). 
\bibitem{Iguchi2015} Y. Iguchi, S. Uemura, K. Ueno, and Y. Onose, Phys. Rev. B {\bf 92}, 184419 (2015).

\bibitem{Seki2016} S. Seki, Y. Okamura, K. Kondou, K. Shibata, M. Kubota, R. Takagi, F. Kagawa,
M. Kawasaki, G. Tatara, Y. Otani, and Y. Tokura, Phys. Rev. B {\bf 93}, 235131 (2016).

\bibitem{Sato2016} T. J. Sato and D. Okuyama, T. Hong, A. Kikkawa, Y. Taguchi, T. H. Arima, 
and Y. Tokura, 
Phys. Rev. B {\bf 94}, 144420 (2016).

\bibitem{Takagi2017} R. Takagi, D. Morikawa, K. Karube, N. Kanazawa, K. Shibata, G. Tatara, Y. Tokunaga,
T. Arima, Y. Taguchi, Y. Tokura, and S. Seki, Phys. Rev. B {\bf 95}, 220406(R) (2017).




\bibitem{24} A. Hrabec, N. A. Porter, A. Wells, M. J. Benitez, G. Burnell, S. McVitie, D. McGrouther, T. A. Moore, and C. H. Marrows, Phys. Rev. B {\bf 90} 020402(R) (2014).


\bibitem{25} R. A. Khan, P. M. Shepley, A. Hrabec, A. W. J. Wells, B. Ocker, C. H. Marrows, and T. A. Moore, Appl. Phys. Lett. {\bf 109} 132404 (2016).


\bibitem{26} C.-F. Pai, M. Mann, A. J. Tan, and G. S. D. Beach, Phys. Rev. B {\bf 93}, 144409 (2016).



\bibitem{27} D. Li, R. Ma, B. Cui, J. Yun, Z. Quan, Y. Zuo, L. Xi, and X. Xu, Appl. Surf. Sci. {\bf 513}, 145768 (2020).




\bibitem{Zhou2020} Y. Zhou, R. Mansell, S. Valencia, F. Kronast, and S. van Dijken, Phys. Rev. B {\bf 101}, 054433 (2020).

\bibitem{Garlow2019} J. A. Garlow, S. D. Pollard, M. Beleggia, T. Dutta, H. Yang, and Y. Zhu, Phys. Rev. Lett. {\bf 122}, 237201 (2019).


\bibitem{Schlotter2018}  S. Schlotter, P. Agrawal, and G. S. D. Beach, Appl. Phys. Lett. {\bf 113}, 092402 (2018).

\bibitem{Bacani2019} M. Ba\'cani, M. A. Marioni, J. Schwenk, and H. J. Hug, Sci. Rep. {\bf 9}, 3114 (2019). 

\bibitem{Agrawal2019} P. Agrawal, F. B\"uttner, I. Lemesh, S. Schlotter, and G. S. D. Beach, Phys. Rev. B {\bf 100}, 104430 (2019).

\bibitem{Zakeri2010} K. Zakeri, Y. Zhang, J. Prokop, T.-H. Chuang, N. Sakr, W. X. Tang, and J. Kirschner, 
Phys. Rev. Lett. {\bf 104}, 137203 (2010);   
K. Zakeri, Y. Zhang, T.-H. Chuang, and J. Kirschner, Phys. Rev. Lett. {\bf 108}, 197205 (2012).  

\bibitem{Korner2015} H. S. K\"orner, J. Stigloher, H. G. Bauer, H. Hata, T. Taniguchi, T. Moriyama, T. Ono, 
and C. H. Back, Phys. Rev. B {\bf 92}, 220413(R) (2015). 

\bibitem{Lee2015a} J. M. Lee, C. Jang, B.-C. Min, S.-W. Lee, K.-J. Lee, and J. Chang, Nano Lett. {\bf 16}, 62 (2015).

\bibitem{Cho2015} J. Cho, N.-H. Kim, S. Lee, J.-S. Kim, R. Lavrijsen, A. Solignac, Y. Yin, D.-S. Han, N. J. J. van Hoof, 
H. J. M. Swagten, B. Koopmans, and C.-Y. You, Nat. Commun. {\bf 6}, 7635 (2015). 

\bibitem{Belmeguenai2015} M. Belmeguenai, J.-P. Adam, Y. Roussign\'e, S. Eimer, T. Devolder, J.-V. Kim, S. M. Cherif, A. Stashkevich, 
and A. Thiaville, Phys. Rev. B {\bf 91}, 180405(R) (2015). 

\bibitem{Di2015} K. Di, V. L. Zhang, H. S. Lim, S. C. Ng, M. H. Kuok, J. Yu, J. Yoon, X. Qiu, and H. Yang, 
Phys. Rev. Lett. {\bf 114}, 047201 (2015). 

\bibitem{Nembach2015} H. T. Nembach, J. M. Shaw, M. Weiler, E. Ju\'e, and T. J. Silva, Nat. Phys. {\bf 11}, 825 (2015). 

\bibitem{Stashkevich2015} A. A. Stashkevich, M. Belmeguenai, Y. Roussign\'e, S. M. Cherif, M. Kostylev, M. Gabor, D. Lacour, C. Tiusan, M. Hehn, Phys. Rev. B {\bf 91}, 214409 (2015).

\bibitem{Chaurasiya2016} A. K. Chaurasiya, C. Banerjee, S. Pan, S. Sahoo, S. Choudhury, J. Sinha, and A. Barman, 
Sci. Rep. {\bf 6}, 32592 (2016). 

\bibitem{Ma2016} X. Ma, G. Yu, X. Li, T. Wang, D. Wu, K. S. Olsson, Z. Chu, K. An, J. Q. Xiao, K. L. 
Wang, and X. Li, Phys. Rev. B {\bf 94}, 180408(R) (2016).

\bibitem{Hrabec2017} A. Hrabec, M. Belmeguenai, A. Stashkevich, S. M. Ch\'erif, S. Rohart, Y. Roussign\'e, 
and A. Thiaville, Appl. Phys. Lett. {\bf 110}, 242402 (2017). 

\bibitem{Robinson2017} R. M. Rowan-Robinson, A. A. Stashkevich, Y. Roussign\'e, M. Belmeguenai, S.-M. Ch\'erif, A. Thiaville, 
T. P. A. Hase, A. T. Hindmarch, and D. Atkinson, Sci. Rep. {\bf 7}, 16835 (2017).  

\bibitem{Ma2017} X. Ma, G. Yu, S. A. Razavi, S. S. Sasaki, X. Li, K. Hao, S. H. Tolbert, K. L. Wang, 
and X. Li, Phys. Rev. Lett. {\bf 119}, 027202 (2017). 

\bibitem{Kim2021} 
N.-H. Kim, Qurat-ul-ain, J. Kim, E. Baek, J.-S. Kim, H.-J. Park, H. Kohno, K.-J. Lee, S. H. Rhim, H.-W. Lee, and C.-Y. You, Phys. Rev. B {\bf 105}, 064403 (2022).


\bibitem{review2023}  
M. Kuepferling, A. Casiraghi, G. Soares, G. Durin, F. Garcia-Sanchez, L. Chen, C. H. Back, C. H. Marrows, S. Tacchi, and G. Carlotti, 
Rev. Mod. Phys. {\bf 95}, 015003 (2023). 



\bibitem{Heide2008} M. Heide, G. Bihlmayer, S. Bl\"ugel, Phys. Rev. B {\bf 78}, 140403(R) (2008).

\bibitem{Katsnelson2010}
 M. I. Katsnelson, Y. O. Kvashnin, V. V. Mazurenko, and A. I. Lichtenstein, Phys. Rev. B {\bf 82}, 100403(R) (2010).

\bibitem{Gayles2015} J. Gayles, F. Freimuth, T. Schena, G. Lani, P. Mavropoulos, 
R. A. Duine, S. Bl\"ugel, J. Sinova, and Y. Mokrousov,  
Phys. Rev. Lett. {\bf 115}, 036602 (2015).

\bibitem{Koretsune2015} T. Koretsune, N. Nagaosa, and R. Arita, Sci. Rep. {\bf 5}, 13302 (2015).

\bibitem{Wakatsuki2015} R. Wakatsuki, M. Ezawa, and N. Nagaosa, Sci. Rep. {\bf 5}, 13638 (2015).


\bibitem{Kikuchi2016} T. Kikuchi, T. Koretsune, R. Arita, and G. Tatara, 
Phys. Rev. Lett. {\bf 116}, 247201 (2016). 

\bibitem{Freimuth2017} 
F. Freimuth, S. Bl\"ugel, and Y. Mokrousov, Phys. Rev. B {\bf 96}, 054403 (2017).

\bibitem{Koretsune2018} 
T.Koretsune, T. Kikuchi, and R. Arita, J. Phys. Soc. Japan {\bf 87}, 041011 (2018).

\bibitem{Ado2018}
I. A. Ado, A. Qaiumzadeh, R. A. Duine, A. Brataas, and M. Titov, Phys. Rev. Lett. {\bf 121},  086802 (2018).

\bibitem{Kataoka1984} M. Kataoka, O. Nakanishi, A. Yanase, and J. Kanamori, 
J. Phys. Soc. Japan {\bf 53}, 3624 (1984).

\bibitem{AGD}  A. A. Abrikosov, L. P. Gor'kov, I. E. Dzyaloshinskii, 
{\it Methods of Quantum Field Theory in Statistical Physics} (Pergamon Press, Oxford). 

\bibitem{Tatara2008} G. Tatara, H. Kohno, and J. Shibata, Phys. Rep. {\bf 468}, 213 (2008). 

\bibitem{Melcher1973} R.~L.~Melcher, Phys. Rev. Lett. {\bf 30}, 125 (1973).

\bibitem{Kataoka1987} M. Kataoka, J. Phys. Soc. Japan {\bf 56}, 3635 (1987).

\bibitem{Udvardi2009} L. Udvardi and L. Szunyogh, Phys. Rev. Lett. {\bf 102}, 207204 (2009). 

\bibitem{TI_review} M. Z. Hasan and C. L. Kane, Rev. Mod. Phys. {\bf 82}, 3045 (2010); 
 X.-L. Qi and S.-C. Zhang, Rev. Mod. Phys. {\bf 83}, 1057 (2011).

\bibitem{TI_spin} Y. Tokura, K. Yasuda, and A. Tsukazaki, Nat. Rev. Phys. {\bf 1}, 126 (2019). 

\bibitem{com1} This information is not lost in our formulation. 
  In fact, anisotropic exchange interaction can be included in 
${\bm M} = (J_\parallel S_x, J_\parallel S_y, J_\perp S_z)$, and thus 
$\varepsilon_{i\alpha} ({\bm M} \times \partial_i {\bm M})^\alpha  
 = \varepsilon_{i\alpha} J_\parallel J_\perp ({\bm S} \times \partial_i {\bm S})^\alpha$. 

\bibitem{Weyl_review} P. Hosur and X. Qi, C. R. Phys. {\bf 14}, 857 (2013); 
N.~P. Armitage, E.~J. Mele, and A. Vishwanath, 
Rev. Mod. Phys. {\bf 90}, 015001 (2018); 
B. Q. Lv, T. Qian, and H. Ding, Rev. Mod. Phys. {\bf 93}, 025002 (2021). 

\bibitem{Kurebayashi2021} D. Kurebayashi, Y. Araki, and K. Nomura, 
J. Phys. Soc. Japan {\bf  90}, 084702 (2021). 

\bibitem{com2} We believe this is allowed in the calculation of the coefficient $D_i^\alpha$, 
in contrast to the calculation of the whole term (chiral anomaly) presented 
in Eq.~(\ref{eq:S_theta_2}). 

\bibitem{HK2021}  H. Kohno, JPSJ News Comments {\bf 18}, 13 (2021). 

\bibitem{Zyuzin2012} A. A. Zyuzin and A. A. Burkov, Phys. Rev. B {\bf 86}, 115133 (2012). 

\bibitem{Liu2013} C.-X. Liu, P. Ye, and X.-L. Qi, Phys. Rev. B {\bf 87}, 235306 (2013). 

\bibitem{Fujikawa_book} K. Fujikawa and H. Suzuki, 
{\it Path Integrals and Quantum Anomalies} (Clarendon Press, Oxford, 2004). 

\bibitem{Adler1969} S. L. Adler, Phys. Rev. {\bf 177}, 2426 (1969); 
 J. S. Bell and R. Jackiw, Nuovo Cimento A {\bf 60}, 47 (1969). 


\bibitem{Rashba2003} E. I. Rashba, Phys. Rev. B {\bf 68}, 241315(R) (2003).

\bibitem{Tokatly2008} I.~V. Tokatly, Phys. Rev. Lett. {\bf 101}, 106601 (2008).

\bibitem{Droghetti2022} A. Droghetti, I. Rungger, A. Rubio, and I.~V. Tokatly, 
Phys. Rev. B {\bf 105}, 024409 (2022).


\bibitem{Ikeda2013} T. Ikeda, Master thesis (Osaka University, March 2013).



\end{thebibliography}
\end{document}